\documentclass[11pt]{article}
\usepackage[margin=1in]{geometry}
\usepackage[T1]{fontenc}
\usepackage{lmodern}
\usepackage{microtype}
\usepackage{amsmath,amssymb,amsthm}
\usepackage{booktabs}
\usepackage[numbers,sort&compress]{natbib}
\usepackage{verbatim}
\usepackage{upquote}
\usepackage[boxed]{algorithm}
\usepackage[noend]{algpseudocode}
\usepackage[hidelinks]{hyperref}

\makeatletter
\newcommand*{\fs@algbox}{\fs@boxed\let\@fs@iftopcapt\iftrue}
\makeatother
\floatstyle{algbox}\restylefloat{algorithm}\floatstyle{plain}

\AddToHook{env/table/begin}{\setlength{\tabcolsep}{4pt}}
\AddToHook{env/claim/begin}{\setlength{\emergencystretch}{2em}}

\newcommand{\E}{\mathbb{E}}
\newcommand{\score}{\operatorname{sc}}

\theoremstyle{plain}
\newtheorem{theorem}{Theorem}[section]
\newtheorem{lemma}[theorem]{Lemma}
\newtheorem{proposition}[theorem]{Proposition}
\newtheorem{corollary}[theorem]{Corollary}
\newtheorem{claim}[theorem]{Claim}
\theoremstyle{definition}
\newtheorem*{modelremark}{Remark}

\title{An Order-Optimal Square-Root Bound for Online Maximin Share Allocation with Known Types}
\author{Zheyu Zhou\\
Department of Mechanical \& Industrial Engineering, University of Toronto\\
\texttt{zheyu.zhou@mail.utoronto.ca}}
\date{}

\begin{document}
\maketitle

\begin{abstract}
An allocator knows the indivisible goods, the number $n$ of agents, and the $k$ possible additive valuations.
Each agent reveals their type on arrival and must immediately receive an irrevocable bundle.
Kulkarni, Mehta, and Shahkar established a deterministic guarantee of $1/k$ of maximin share (MMS) and an upper bound of $2/(\sqrt{k}-2)$ for $k>4$.
We determine the optimal order $\Theta(k^{-1/2})$ by constructing a deterministic polynomial-time algorithm, \emph{Single-or-Select}.
On every arrival sequence, it gives each agent a fraction $(1/\sqrt{2}-o(1))/\sqrt{k}$ of their truncated proportional share.
This share dominates MMS and is the largest target $y$ for which the item values capped at $y$ sum to at least $ny$.
By specializing the construction of Kulkarni et al. to two arrivals and counting its types exactly, we sharpen the upper bound to $(\sqrt{2}+o(1))/\sqrt{k}$ even for binary valuations.
For two arrivals with binary valuations and every $k\ge2$, the optimal fraction is $1/p$, where $p$ is the largest integer with $1+\binom{2p}{2}\le k$. In particular, up to six binary types permit full maximin share.
For two arbitrary additive types, a different polynomial-time algorithm guarantees $4/7$ of maximin share for any number of agents.
This ratio is tight for the algorithm. With three arrivals and two types, the optimal deterministic fraction is exactly $9/10$.
For randomized policies with $k$ known additive types, the optimal MMS fraction guaranteed to all agents with probability at least $1-\delta$ on every fixed arrival sequence is $\Theta(\max\{k^{-1/2},\sqrt{\delta}\})$ for $0\le\delta<1$.

\end{abstract}

\par

\section{Introduction}
Consider an allocator facing two arrivals and four indivisible goods. The first agent values every good at one. She can divide the goods into two pairs, so her maximin share (MMS) is two. Suppose the list of possible valuations also includes, for each pair of goods, a type that values only the goods in that pair, at one apiece. If the allocator gives the first agent two goods, the second agent may value just those two and receive nothing, although her MMS is one. Giving the first agent a single good protects the second at the cost of limiting the first to half her share. This is the smallest two-agent instance in the construction of Kulkarni, Mehta, and Shahkar~\cite[Sec.~4.2]{KMS} that forces a loss of MMS.

How much of each agent's share can an allocator guarantee when it knows the $k$ possible valuations and the total number of arrivals, but not how often each type will occur?
Under adversarial arrivals, Kulkarni et al. give a deterministic policy that guarantees every agent at least $1/k$ of her MMS~\cite[full version, Theorem~3.1, p.~8]{KMS}.
For $k>4$, the optimal worst-case MMS fraction is at most $2/(\sqrt{k}-2)$, even for binary valuations~\cite[full version, Theorem~3.3, p.~9]{KMS}.
This upper bound rules out an $\omega(1/\sqrt{k})$ guarantee and leaves open whether $c/\sqrt{k}$ is attainable for an absolute constant $c>0$.
Our deterministic polynomial-time algorithm attains such a guarantee for any number of agents (Theorem~\ref{thm:general}).

\subsection{Results and techniques}
\paragraph{An order-optimal deterministic algorithm.}
\emph{Single-or-Select} (Algorithm~\ref{alg:general}) guarantees each agent a fraction $1/(2h)$ of her truncated proportional share (TPS), where $h=\sqrt{k/2}+O(1)$ (Theorem~\ref{thm:general}). Since TPS is at least MMS, this matches the order of the MMS upper bound.
Our starting point is \emph{Single-or-Sample}, introduced in the September~2026 preprint of Chawla et al.~\cite[Algorithm~2 and Lemma~4.3]{SOS}. Under their safe strategy for additive valuations, an agent takes a single good if one meets her target. Otherwise, she orders the remaining goods by value and receives one uniformly random good from each full bucket. Sorting gives her a lower bound on the value of her bundle regardless of which good is drawn from each bucket.

We use this freedom to choose the goods deterministically while preserving value for every possible future type. We adapt their bucket width rule by deducting earlier empty rounds as well as singleton rounds from the initial number of agents. On the first sampling round, we bound each type's loss using products of weights over ordered pairs of goods from distinct buckets. Under independent uniform draws, the expected sum is at most $1/h^2$ per type and hence at most $k/h^2$ over all types. A width multiplier $h$ of order $\sqrt{k}$ thus keeps this bound constant as $k$ grows. We combine the pair estimate with a product that tracks each type's cumulative weight loss, then use conditional expectations~\cite{Raghavan} to select one good per full bucket.

The pair construction in the opening example extends to $2p$ goods and $1+\binom{2p}{2}$ types. Counting these types in the construction of Kulkarni et al. gives the upper coefficient $\sqrt{2}$. For finite $k$, Corollary~\ref{cor:finite-envelope} compares the three explicit guarantees and gives the ranges where each is strongest.

If the arrival sequence is fixed before the policy makes its random choices, the optimal MMS fraction guaranteed simultaneously to all agents with failure probability at most $\delta$ is $\Theta(\max\{k^{-1/2},\sqrt{\delta}\})$ (Proposition~\ref{prop:oblivious}). Thus $\delta=O(1/k)$ preserves the deterministic order, and a larger asymptotic order is attainable precisely when $k\delta\to\infty$.

\paragraph{An exact result for binary valuations.}
For two arrivals and binary valuations, full MMS is possible for $k\le6$, and we match the pair construction at every $k\ge7$ (Theorem~\ref{thm:binary}).
For $k\le6$, we pair goods whenever every other type approving one also approves the other. Giving the less widely approved good to the first agent preserves at least half the approvals in each pair for every possible second type. The remaining patterns form an antichain on at most five coordinates. A finite enumeration proves that these goods can complete an MMS allocation. For $k\ge7$, the first agent needs a smaller fraction of her approved goods. For each possible second type, we designate disjoint pairs of its approved goods. Whenever both goods of a pair are available to the first agent, we join them by an edge. An independent set then gives the first agent goods while leaving at least one endpoint of each designated pair for the second agent. A bound on the edge count guarantees an independent set of the required size. The proof finds one in polynomial time. In particular, the half-share obstruction in the opening example gives the worst case over all seven-type binary instances with two arrivals.

\paragraph{Two types with arbitrary item values.}
For two arbitrary additive types, Section~\ref{sec:two-overview} proves that a candidate-bundle policy guarantees $4/7$ of MMS for any number of agents (Theorem~\ref{thm:two}), that this fraction is tight for the policy (Proposition~\ref{prop:policy}), and that the optimal three-agent guarantee is $9/10$ (Theorem~\ref{thm:three}). Hence the optimum for any number of agents lies between $4/7$ and $9/10$. Table~\ref{tab:bounds} summarizes the deterministic bounds.

\paragraph{Formal verification.}
The accompanying Lean~4.19 formalization~\cite{ZhouLean} covers the numbered results in the paper, including the deterministic $\Theta(k^{-1/2})$ guarantee and the randomized guarantee of Proposition~\ref{prop:oblivious}. It also verifies the adaptive-arrival equivalence and the $1/k$ guarantee of Kulkarni et al.~\cite{KMS}. Lean's kernel checks the count of $7{,}581$ antichains and their balanced selections in Lemma~\ref{lem:five-patterns}, as well as the finite comparisons in Corollary~\ref{cor:finite-envelope}. The formalization is publicly available in the cited repository, which maps each result to its proof.

\begin{table}[tb]
\caption{Worst-case deterministic fractions of original MMS. In the rows marked \emph{Any $n$}, the lower bound holds for each known $n$ and the upper bound applies to the worst case over $n$. In the binary row, $p$ is the largest integer with $1+\binom{2p}{2}\le k$. All three lower bounds use polynomial-time policies.}
\label{tab:bounds}
\begin{tabular}{@{}llcll@{}}
\toprule
Valuations & Number of agents & Lower bound & Upper bound & Result \\
\midrule
$k$ additive types & Any $n$ & $(1/\sqrt{2}-o(1))/\sqrt{k}$ & $(\sqrt{2}+o(1))/\sqrt{k}$ & Thm.~\ref{thm:general}, Prop.~\ref{prop:upper} \\
$k\ge2$ binary types & $n=2$ & $1/p$ & $1/p$ & Thm.~\ref{thm:binary} \\
Two additive types & Any $n$ & $4/7$ & $9/10$ & Thms.~\ref{thm:two}, \ref{thm:three} \\
\bottomrule
\end{tabular}
\end{table}

\subsection{Related work}
Budish~\cite{Budish} introduced the maximin share as a benchmark for indivisible goods. Even with all agents present, an allocation giving everyone full MMS can fail to exist. For additive valuations, Kurokawa, Procaccia, and Wang~\cite{KPW} proved that $2/3$ is always attainable, and Akrami and Garg~\cite{AG} gave a guarantee strictly above $3/4$. With two known types and known multiplicities, Shahkar and Garg~\cite{SG} give a $4/5$ guarantee. Ezra and Garbuz~\cite[Theorem~2]{EG} exhibit a four-agent, two-type instance whose optimum is at most $20/21$. For three agents, Feige and Norkin~\cite[Theorems~12 and~17]{FN} give an offline $11/12$ guarantee and study the additional restriction that one designated agent receive a whole part of her MMS partition. Section~\ref{sec:two-overview} uses the latter result to obtain the online $9/10$ guarantee and their matrix to prove a matching upper bound.

Aleksandrov and Walsh~\cite{AW} survey online fair division models in which the available information and the timing of decisions vary. In the item-arrival model of Zhou, Bai, and Wu~\cite{ZBW}, a fixed group of agents faces a stream of goods. With normalized additive valuations they obtain the tight guarantee $1/2$ for two agents. Here the allocator knows the goods at the outset and learns which of the known valuations to serve as agents arrive. Kulkarni et al.~\cite{KMS} introduced this known-type formulation and proved the $1/k$ guarantee and the pair obstruction that we use in Proposition~\ref{prop:upper}.

Without advance knowledge of the type list, Chawla et al.~\cite[Theorem~4.1 and Corollary~4.2]{SOS} obtain a simultaneous guarantee for submodular agents. On every input fixed before the random choices, all agents following their safe strategies receive $1/24$ of MMS with probability at least $0.423$. Their Lemma~4.3 scales the bucket width by $n$ minus the number of earlier singleton rounds. Our reference total $N=r+s$ (remaining agents plus earlier samples) also deducts zero-TPS rounds from $n$.
For additive goods with no advance knowledge of the type list, Li, Wang, and Wu~\cite[Theorem~E.2]{LWW} guarantee every agent a fraction $x$ of MMS simultaneously with probability at least $1-85x^2$ for $0<x\le1/15$. Appendix~E of their preprint also gives an expected minimum fraction of at least $1/10$ under a different parameter choice (Theorem~E.1) and, for every randomized policy and even $n$, a fixed binary instance with expected minimum at most $1/2+O(1/\sqrt n)$ (Theorem~E.3).

For chores with $k$ known additive cost types, Li, Lin, Miao, and Wu~\cite[Sec.~4]{LLMW} construct a partition in advance with approximation factor $O(\min\{\log k,\log(kn)/\log\log(kn)\})$. Their universal-residual algorithm~\cite[Sec.~5 and Appendix~E]{LLMW} handles $k\le n$ binary cost types. After a type arrives, it chooses a bundle that keeps the remaining chores feasible for every type. Conditional expectations on a potential over types derandomize a uniform choice of a fixed-size subset. The resulting cost is at most three times MMS. They prove a lower bound of two for some constant $k$, exact MMS for two types, and a $3/2$ lower bound for seven types. For goods we allow arbitrary additive values and any $k$. With TPS-normalized weights, our pair estimate for the first sample permits a width parameter of $\sqrt{k/2}+O(1)$, while a product protects all types through later samples (Section~\ref{sec:general}). Alon et al.~\cite{AAABN} use a related potential over uncovered elements to derandomize online set cover when the ground set and set system are known.

\section{Model and benchmarks}\label{sec:model}
An instance $I=(n,M,v_1,\ldots,v_k)$ specifies $n\ge1$ agents, a set $M$ of $m$ indivisible goods, and $k\ge1$ possible types. Write $[k]=\{1,\ldots,k\}$. Each type $t\in[k]$ has a nonnegative additive valuation $v_t\colon 2^M\to\mathbb R_{\ge0}$. Write $v_t(g)=v_t(\{g\})$ for each good $g\in M$. Then $v_t(S)=\sum_{g\in S}v_t(g)$ for every bundle $S\subseteq M$. Let $\mathcal I_{k,n}$ be the set of such instances. The allocator knows $I$ in advance. The agents arrive in an unknown sequence of types $\mathbf t=(t_1,\ldots,t_n)\in[k]^n$. A type may occur any number of times, including zero. Upon learning $t_j$, the allocator immediately and irrevocably gives agent $j$ a bundle of remaining goods. Bundles may be empty. Since values are nonnegative, we give the last agent all remaining goods.

We compute each agent's share using the original set $M$ and all $n$ agents, regardless of when she arrives. The \emph{original maximin share} (MMS) of a type-$t$ agent is
\begin{equation}\label{eq:mms}
 \mu_t=\max_{U_1\sqcup\cdots\sqcup U_n=M}\ \min_{i\in[n]}v_t(U_i).
\end{equation}
This is the value she can guarantee herself by partitioning $M$ into $n$ possibly empty bundles and choosing last.

Her truncated proportional share (TPS) is
\begin{equation}\label{eq:tps}
 d_t = \max\biggl\{y\ge0:\ \sum_{g\in M}\min\{v_t(g),y\}\ge ny\biggr\}.
\end{equation}
The proportional share $v_t(M)/n$ counts the full value of every good. At a target of $y$ per bundle, a single good can supply at most $y$ toward one bundle's target. Its excess cannot help fill a second bundle.

Babaioff, Ezra, and Feige~\cite{BEF} introduced TPS and proved the share bounds below. Iterating their recursion gives the minimum in~\eqref{eq:tps-closed}.
\begin{lemma}[Truncated proportional share]\label{lem:tps}
Let $x_1\ge\cdots\ge x_m$ be type $t$'s item values and let $S_c=\sum_{i=c+1}^m x_i$. The maximum in~\eqref{eq:tps} is attained, and
\begin{equation}\label{eq:tps-closed}
 d_t=\min_{0\le c\le\min(m,n-1)}\frac{S_c}{n-c},
 \qquad \sum_{g\in M}\min\{v_t(g),d_t\}=nd_t .
\end{equation}
Moreover, $\mu_t\le d_t\le v_t(M)/n$.
\end{lemma}
\begin{proof}
For $y\ge0$, the capped sum is $\min_{0\le c\le m}(cy+S_c)$, attained at $c=|\{i:x_i>y\}|$. Hence $y$ is feasible exactly when $(n-c)y\le S_c$ for all $0\le c\le\min(m,n-1)$. The smallest ratio in~\eqref{eq:tps-closed} is therefore the largest feasible $y$. At a minimizing $c$, the capped sum at $d_t$ is at most $cd_t+S_c=nd_t$ and at least $nd_t$ by feasibility. The $c=0$ term in the minimum gives the proportional upper bound. After capping at $\mu_t$, each part of an MMS partition either contains a good that supplies $\mu_t$ on its own or retains its full value. Summing over the $n$ parts gives at least $n\mu_t$, so $\mu_t$ is feasible.
\end{proof}

\paragraph{Policy guarantees.}
On a fixed instance $I$, a deterministic policy $\pi=(\pi_1,\ldots,\pi_n)$ consists of maps $\pi_j\colon[k]^j\to 2^M$. Write $\mathbf t_{\le j}=(t_1,\ldots,t_j)$. A policy is feasible if, on every sequence $\mathbf t\in[k]^n$, the bundles $\pi_j(\mathbf t_{\le j})$ are pairwise disjoint and have union $M$.
For a target vector $\tau\in\mathbb R_{\ge0}^k$, a feasible policy $\pi$ \emph{guarantees $\tau$ on $I$} if $v_{t_j}(\pi_j(\mathbf t_{\le j}))\ge\tau_{t_j}$ for every $\mathbf t\in[k]^n$ and every $j\in[n]$.
The fraction of original MMS that all agents receive, capped at one, is
\[
 \rho(\pi,I,\mathbf t)
 =\min\left(\{1\}\cup
 \left\{\frac{v_{t_j}(\pi_j(\mathbf t_{\le j}))}{\mu_{t_j}}:
                 j\in[n],\ \mu_{t_j}>0\right\}\right).
\]
The optimal worst-case guarantees for a known number $n$ of agents and for any $n$, respectively, are
\[
 \alpha_{k,n}=\inf_{I\in\mathcal I_{k,n}}\ \max_{\pi\text{ feasible on }I}\
                      \min_{\mathbf t\in[k]^n}\rho(\pi,I,\mathbf t),
 \qquad \alpha_k=\inf_{n\ge1}\alpha_{k,n}.
\]
There are finitely many prefixes and bundle choices on each $I$, so the policy maximum exists. Replacing $\mu_{t_j}$ by $d_{t_j}$ in $\rho$ defines $\alpha_{k,n}^{\mathrm{TPS}}$ and $\alpha_k^{\mathrm{TPS}}$. Lemma~\ref{lem:tps} gives
\[
 \alpha_{k,n}^{\mathrm{TPS}}\le\alpha_{k,n},\qquad
 \alpha_k^{\mathrm{TPS}}\le\alpha_k .
\]

\begin{lemma}[Adding one agent]\label{lem:padding}
For each $k,n\ge1$, $\alpha_{k,n+1}\le\alpha_{k,n}$ and
$\alpha^{\mathrm{TPS}}_{k,n+1}\le\alpha^{\mathrm{TPS}}_{k,n}$.
Consequently $\alpha_k=\lim_{n\to\infty}\alpha_{k,n}$, and the same identity holds for TPS.
\end{lemma}
\begin{proof}
Add one agent and a good $z$ with $v_t(z)=\mu_t$ for every type $t$. Appending $\{z\}$ to each type's original MMS partition preserves its minimum. Conversely, deleting the part containing $z$ and redistributing its old goods turns any augmented partition into an $n$-part partition with at least the same minimum. For TPS, set $v_t(z)=d_t$. The new capped sum at $d_t$ is $(n+1)d_t$ by~\eqref{eq:tps-closed}. At $y>d_t$ the old sum is less than $ny$ and $z$ adds at most $y$. Thus each construction preserves the corresponding benchmark.

Submit each real arrival to a policy for the augmented instance. If its bundle contains $z$, reserve its old goods and submit the same type again. Each real agent receives the returned bundle that omits $z$. At most one extra call is needed, so each agent receives the virtual guarantee within $n+1$ calls. Give leftover old goods to the last agent. Taking infima over the original instances gives both inequalities. Each bounded nonincreasing sequence converges to its infimum.
\end{proof}

\paragraph{Randomized guarantees for fixed sequences.}
An oblivious adversary fixes the type sequence before the policy makes its random choices. For $0\le\delta<1$, the optimal simultaneous MMS fraction $\alpha_k^{\mathrm{obl}}(\delta)$ is the supremum of $x\in[0,1]$ such that, for every $n\ge1$, each $I\in\mathcal I_{k,n}$ admits one distribution $\mathcal D$ on feasible deterministic policies satisfying
\[
 \Pr_{\pi\sim\mathcal D}\!\left[\rho(\pi,I,\mathbf t)\ge x\right]\ge1-\delta
 \qquad\text{for every }\mathbf t\in[k]^n .
\]
\begin{modelremark}[Probability-one guarantees]
If any policy in the finite support of $\mathcal D$ fails on a sequence, that sequence has positive failure probability. Thus at $\delta=0$ the randomized and deterministic optima agree on every instance, giving $\alpha_{k,n}$ for fixed $n$ and $\alpha_k^{\mathrm{obl}}(0)=\alpha_k$.
\end{modelremark}

\begin{proposition}[Simultaneous MMS under oblivious arrivals]\label{prop:oblivious}
Uniformly for $k\ge1$ and $0\le\delta<1$,
\[
 \alpha_k^{\mathrm{obl}}(\delta)
   =\Theta\!\left(\max\{k^{-1/2},\sqrt{\delta}\}\right).
\]
\end{proposition}
The \hyperref[proof:oblivious]{proof at the end of Section~\ref*{sec:binary}} combines Theorem~\ref{thm:general} with a randomized bucket policy for the lower bound and averages over the pairs of Proposition~\ref{prop:upper} for the upper bound.

\begin{modelremark}[Adaptive arrivals]
Every play of a deterministic policy against an adaptive adversary follows a fixed type sequence. Fix an instance and target, and work backwards from the last allocation. A history is winning if every next type admits a bundle meeting its target and leaving a winning history. At any losing history the adversary can choose a type for which every bundle fails immediately or leaves another losing history. Observing each allocation thus lets the adversary force failure on every play whenever the deterministic target is unattainable. The optimum is the same for every $\delta<1$.
\end{modelremark}

\paragraph{Computational model.}
The policy optima impose no computational restriction. For algorithmic bounds, $L$ is the total binary length of $n,k,m$ and the $km$ explicitly listed rational item values. A policy is polynomial-time if preprocessing and each arrival take a number of bit operations polynomial in $L$. If $M=\varnothing$, every benchmark is zero and every policy we implement returns $\varnothing$ on each arrival. For $m>0$, the input lists at least one value per type. Sorting and taking suffix sums in~\eqref{eq:tps-closed} compute TPS exactly. Computing MMS is NP-hard even for two agents with integer values, since deciding whether $\mu_t=v_t(M)/2$ is \textsc{Partition}. For fixed $\varepsilon>0$, Woeginger's PTAS~\cite{Woeginger} finds a partition with minimum value at least $(1-\varepsilon)\mu_t$ whenever $\mu_t>0$.

\section{A deterministic square-root guarantee}\label{sec:general}
We analyze \emph{Single-or-Select}, our deterministic online allocation rule for any number of known types. Its guarantee is relative to the original TPS and therefore applies to the original MMS as well.

\begin{theorem}\label{thm:general}
For $k\ge1$, let $h=4(53+\lceil\sqrt{k/32}\rceil)\le\sqrt{k/2}+216$.
For every $n\ge1$, every $I\in\mathcal I_{k,n}$, and every type sequence
$\mathbf t\in[k]^n$, Algorithm~\ref{alg:general} gives each agent $j$ a
bundle $T_j$ with $v_{t_j}(T_j)\ge d_{t_j}/(2h)$. Consequently,
\[
 \alpha_k\ge\alpha_k^{\mathrm{TPS}}\ge\frac1{2h}
 \ge\frac1{\sqrt{2k}+432},\qquad
 \liminf_{k\to\infty}\sqrt{k}\,\alpha_k\ge
 \liminf_{k\to\infty}\sqrt{k}\,\alpha_k^{\mathrm{TPS}}\ge\frac1{\sqrt2}.
\]
On rational inputs of bit length $L$, the algorithm uses
$O(k(m+1)\log(m+2))$ operations in preprocessing and
$O(k(m+1)+m\log(m+2))$ per arrival, on numbers of $O(k(m+1)L)$ bits.
\end{theorem}

For finite $k$, $1/(2h)$ strictly improves on the $1/k$ MMS guarantee
of Kulkarni et al.~\cite[Theorem~3.1]{KMS} exactly when $k\ge457$. It is the strongest
of the three bounds in Corollary~\ref{cor:finite-envelope} for $k\ge1641$,
with an earlier tie for $1561\le k\le1568$.

\paragraph{Overview and choice of parameters.}
For each type $t$ with $d_t>0$, divide its value for each good by $d_t$
and cap the resulting weight at one. These initial weights sum to $n$.
A bundle of weight at least $\beta=1/(2h)$ for that type has value at
least $\beta d_t$. A nonfinal agent with zero TPS receives the empty bundle.
Following the single-good and bucketing rule of Chawla et al.~\cite[Algorithm~2]{SOS},
give a nonfinal agent with positive TPS a single good if its weight reaches
$\beta$. If no good does, sort the remaining goods by her weights and split
the list into consecutive \emph{buckets} of width $H$, a positive integer.
A full bucket contains $H$ goods. Only the last bucket may be shorter.
A \emph{candidate sample} consists of exactly one good from each full bucket
and no other goods. Sorting gives every candidate the required weight for
the current agent. On a sampling round the algorithm chooses among these
bundles to protect future agents, who may have any type, including hers.

Let $r$ count the agents still to be served and $s$ the sampling rounds
already completed. The reference total $N=r+s$ is $n$ minus the number of
earlier nonsampling rounds. Before the first sample, each positive-TPS type
still has weight $N$. Width $hN$ therefore gives the current agent at least
$1/h-\beta$ by the sorted-bucket bound below. Balancing this with the
singleton guarantee $\beta$ gives $\beta=1/(2h)$.

To see what fixes $h$, suppose only two agents remain at the first sample.
The last agent may need any type's remaining weight, so the sample must
leave at least $\beta$ out of its two units. A loss above $2-\beta$ requires
weight from at least two buckets. If one good is drawn independently and
uniformly from each full bucket, the sum of weight products over unordered pairs
has mean at most $1/(2h^2)$ for each type. Such a loss forces the sum above
$1-\beta$. By Markov's inequality and a union bound, the probability that
some type with positive TPS loses more than $2-\beta$ is at most
$k/[2(1-\beta)h^2]$. This explains $h\sim\sqrt{k/2}$. The scale $1/h$ matches the leading upper
bound in Proposition~\ref{prop:upper}. Halving it to balance singletons and
samples accounts for the factor of two between the proved coefficients.

For each positive-TPS type, the \emph{deficit} is $N$ minus its remaining weight.
On an empty or singleton round, $N$ drops by one and at most one unit of
weight is removed. Trimming to the new $N$ therefore cannot increase the
deficit. On a sampling round $N$ stays fixed. A product bounds the deficit
across these rounds. The bound grows by
$3/4$ per sample after the second. Later the proof guarantees only
$N/4+\beta$ remaining weight before a nonfinal arrival, so the algorithm
uses buckets four times narrower to maintain the same target. The product
alone gives a weaker first-sample estimate, as shown in
Section~\ref{sec:first-sample}.

\subsection{Weights and the allocation rule}
Let $K_+=\{t\in[k]:d_t>0\}$ be the set of \emph{protected types}.
At the start of round $j$, the remaining goods are
$R_j=M\setminus\bigcup_{i<j}T_i$. Agent $j$ has type $t_j$, whereas
$t$ denotes any type in $K_+$. Write $w_t^{(j)}(g)$ for the current weight
of $g\in R_j$ and $w_t^{(j)}(G)=\sum_{g\in G}w_t^{(j)}(g)$ for a bundle
$G\subseteq R_j$. Initially $w_t^{(1)}(g)=\min\{1,v_t(g)/d_t\}$.
Thus $d_t w_t^{(1)}(g)\le v_t(g)$, and~\eqref{eq:tps-closed} gives
$w_t^{(1)}(M)=n$. When a round is fixed, we use $R,w_t,r,s,N,P_t$ for its
current state. In comparisons between rounds we use $R_j,w_t^{(j)},r_j,s_j,
N_j,P_t^{(j)}$, with $r_j=n-j+1$ and $N_j=r_j+s_j$.

On every sampling round, Algorithm~\ref{alg:general} sorts $R$ anew for
the current type $t_j$. With full buckets $B_1,\ldots,B_q$, each candidate
sample has the form $G=\{g_1,\ldots,g_q\}$ with $g_\ell\in B_\ell$.
The agent thus receives exactly $q=\lfloor|R|/H\rfloor$ goods.
Both widths $hN$ and $bN=(h/4)N$ are integers
because $4$ divides $h$. For a candidate $G$, $Q_t(G)$ sums the products
of type-$t$ weights over ordered pairs of distinct selected goods. Two goods
of weight one contribute two to this sum and can exhaust the last agent's reserve.
The updated product $P_t^+(G)$ tracks losses to type $t$ across samples.

\begin{algorithm}[H]
\caption{\emph{Single-or-Select} for known valuation types}\label{alg:general}
\linespread{0.97}\small
\begin{algorithmic}[1]
\State If $M=\varnothing$, return the policy that gives $\varnothing$ on every arrival.
\State Fix an order on $M$. Set $h\gets4(53+\lceil\sqrt{k/32}\rceil)$, $b\gets h/4$, $\beta\gets1/(2h)$.
\State Compute each $d_t$ by~\eqref{eq:tps-closed}. Set $K_+\gets\{t:d_t>0\}$, $R\gets M$, $s\gets0$.
\State For $t\in K_+$, set $P_t\gets1$ and $w_t(g)\gets\min\{1,v_t(g)/d_t\}$ for every $g\in M$.
\Statex For $q$ full buckets, define the following polynomials on weight arrays $x=(x_{t\ell})\in[0,1]^{K_+\times[q]}$.
\Statex \hspace{\algorithmicindent}$\displaystyle Q_t(x)=\sum_{\ell\ne\ell'}x_{t\ell}x_{t\ell'},
 \qquad P_t^+(x)=P_t\prod_{\ell=1}^q(1+(h-1)x_{t\ell}),$
\Statex \hspace{\algorithmicindent}$\displaystyle \Phi_1(x)=\sum_{t\in K_+}\bigl(Q_t(x)+128P_t^+(x)/h^3\bigr),
 \qquad \Phi_2(x)=\sum_{t\in K_+}P_t^+(x).$
\Statex For a candidate $G=\{g_1,\ldots,g_q\}$ with $g_\ell\in B_\ell$, set $x_{t\ell}(G)=w_t(g_\ell)$ and write $f(G)=f(x(G))$ for any function $f$ of $x$.
\For{$j=1,\ldots,n$}
  \State Observe $t_j$. Set $r\gets n-j+1$, $N\gets r+s$.
  \State \textbf{if} $r=1$ \textbf{then} assign $R$ to agent $j$ and stop.
  \If{$t_j\notin K_+$} \State $T_j\gets\varnothing$.
  \ElsIf{some $g\in R$ has $w_{t_j}(g)\ge\beta$}
    \State $T_j\gets\{g\}$ for the first such $g$ in the fixed order.
  \Else
    \State Use $(H,\Phi)=(hN,\Phi_1)$ if $s=0$, and $(bN,\Phi_2)$ otherwise.
    \State Sort $R$ by decreasing $w_{t_j}(g)$, breaking ties in the fixed order.
    \State Split the list into buckets of width $H$. Pass only the full buckets $B_1,\ldots,B_q$ to \textsc{Select}.
    \State $T_j\gets\Call{Select}{B_1,\ldots,B_q,\Phi}$.
    \State Set $P_t\gets P_t^+(T_j)$ for each $t\in K_+$ and $s\gets s+1$.
  \EndIf
  \State Assign $T_j$ to agent $j$ and set $R\gets R\setminus T_j$.
  \State If the zero-TPS or singleton branch was used, call $\Call{Trim}{R,N-1}$.
\EndFor
\Procedure{Select}{$B_1,\ldots,B_q,\Phi$}
  \State Set $x_{t\ell}\gets w_t(B_\ell)/|B_\ell|$ for every $t\in K_+$ and $\ell\in[q]$.
  \For{$\ell=1,\ldots,q$}
    \State For each $g\in B_\ell$, evaluate $\Phi(x)$ with column $\ell$ replaced by $(w_t(g))_{t\in K_+}$.
    \State Choose the first minimizing good in the fixed order and set column $\ell$ to its weight vector.
  \EndFor
  \State \Return the set of chosen goods.
\EndProcedure
\Procedure{Trim}{$R,\mathit{cap}$}
  \For{$t\in K_+$}
    \State $\mathit{excess}\gets\max\{0,w_t(R)-\mathit{cap}\}$.
    \For{successive $g\in R$ in the fixed order while $\mathit{excess}>0$}
      \State $\delta\gets\min\{\mathit{excess},w_t(g)\}$. Subtract $\delta$ from $w_t(g)$ and from $\mathit{excess}$.
    \EndFor
  \EndFor
\EndProcedure
\end{algorithmic}
\end{algorithm}

Each good $g$ in a sample contributes a factor at least $h^{w_t(g)}$ to
$P_t$, so the product bounds $h$ raised to the cumulative sampled weight.
The coefficient $128/h^3$ in the first objective $\Phi_1$ ensures that
keeping $\Phi_1$ below two leaves a product budget of $h^3/64$.
The divisor $64>e^4$ covers the growth bound for one later sample.
The later objective $\Phi_2$ uses the products alone.

\subsection{What sorting and trimming preserve}
The following additive bound uses the consecutive-bucket argument in the
proof of Chawla et al.~\cite[Lemma~4.3, Eq.~(4)]{SOS}.
\begin{lemma}[Weight of a candidate from sorted buckets]\label{lem:payment}
Fix a round $j$ with $t_j\in K_+$ and sort $R$ by nonincreasing $w_{t_j}$.
For any positive integer width $H$, every candidate sample $G$ satisfies
\[
 w_{t_j}(G)\ge \frac{w_{t_j}(R)}H-\max_{g\in R}w_{t_j}(g),
\]
where the maximum is zero for $R=\varnothing$.
\end{lemma}
\begin{proof}
The claim is immediate when $R$ is empty. Otherwise the first bucket weighs
at most $H\max_{g\in R}w_{t_j}(g)$. Each subsequent bucket has at most $H$
goods, all no heavier than the good chosen from the preceding full bucket.
Together these subsequent buckets weigh at most $H w_{t_j}(G)$.
\end{proof}

To track the goods left for future agents, define the \emph{deficit} of
type $t$ at the start of round $j$ by
\begin{equation}\label{eq:deficit}
 D_t^{(j)}=N_j-w_t^{(j)}(R_j).
\end{equation}
Write $D_t$ within a fixed round. Trimming enforces the upper bound
$w_t(R)\le N$ needed to bound how much sampling can remove.

\begin{lemma}[State invariants]\label{lem:trim}
At the start of every round, each $t\in K_+$ satisfies
\[
 0\le w_t(R)\le N,\qquad h^{D_t}\le P_t,\qquad
 d_t w_t(G)\le v_t(G)\quad\text{for every }G\subseteq R.
\]
For each such type, $D_t^{(1)}=0$. In round $j<n$, the zero-TPS and
singleton branches give $D_t^{(j+1)}\le D_t^{(j)}$, and the sampling
branch gives $D_t^{(j+1)}=D_t^{(j)}+w_t^{(j)}(T_j)$.
\end{lemma}
\begin{proof}
Initially $w_t^{(1)}(M)=n$, $D_t^{(1)}=0$ and $P_t^{(1)}=1$.
Trimming only lowers nonnegative weights, and sampling leaves all
retained weights unchanged. Thus every retained good satisfies
$d_t w_t^{(j)}(g)\le d_t w_t^{(1)}(g)\le v_t(g)$. Summing gives the bundle bound.
On the zero-TPS or singleton branch of a nonfinal round, $N_{j+1}=N_j-1$ and at most one
unit of weight is allocated. Trimming to the new total gives
\begin{equation}\label{eq:singleton}
 D_t^{(j+1)}=\max\{0,D_t^{(j)}-1+w_t^{(j)}(T_j)\}\le D_t^{(j)} .
\end{equation}
The inequality uses $D_t^{(j)}\ge0$ from the induction hypothesis.
In a sampling round, $r$ drops by one and $s$ rises by one, so
$N_{j+1}=N_j$. The only weight removed is $w_t^{(j)}(T_j)$.
These updates preserve $0\le w_t(R)\le N$. Convexity gives
$h^x\le1+(h-1)x$ for $0\le x\le1$. Multiplying this inequality over
the selected goods shows that the update to $P_t$ covers the increase
in $h^{D_t}$, while nonsampling rounds leave $P_t$ unchanged.
\end{proof}

\subsection{Deterministic selection}
Fix a sampling round, its starting state, and its full buckets.
For the analysis, let $\widetilde g_\ell$ be drawn uniformly from $B_\ell$,
independently across buckets, and put $\widetilde G=\{\widetilde g_1,\ldots,
\widetilde g_q\}$. Expectations below refer to this experiment in the
specified round. The algorithm itself makes a deterministic choice.

\begin{lemma}[Conditional expectations]\label{lem:select}
Let $\Phi\colon[0,1]^{K_+\times[q]}\to\mathbb R$ be multi-affine in the
bucket columns, meaning that it is affine in each column with the others fixed. Then
\textsc{Select} returns a candidate sample $G$ with $\Phi(G)\le\E[\Phi(\widetilde G)]$.
The objectives $\Phi_1$ and $\Phi_2$ in Algorithm~\ref{alg:general} are
both multi-affine. For either objective, \textsc{Select} uses $O(k(m+1))$ operations.
\end{lemma}
\begin{proof}
Write $\mathbf w(g)=(w_t(g))_{t\in K_+}$ for the weight vector of good $g$,
and let $\bar{\mathbf w}_\ell$ be the mean of these vectors over $B_\ell$.
Fix all columns except column $\ell$, and write $F(u)$ for the value of
$\Phi$ when that column is $u$. Since $F$ is affine,
\begin{equation*}
 \frac{1}{|B_\ell|}\sum_{g\in B_\ell} F(\mathbf w(g))
 = F\!\left(\frac{1}{|B_\ell|}\sum_{g\in B_\ell}\mathbf w(g)\right)
 = F(\bar{\mathbf w}_\ell).
 \tag{*}\label{eq:select-bucket-mean}
\end{equation*}

Suppose \textsc{Select} has chosen $g_1,\ldots,g_{\ell-1}$. Conditional
on the random sample making these same choices, the remaining draws are
still independent and uniform. Applying~\eqref{eq:select-bucket-mean} to one
bucket at a time, from $B_q$ back to $B_\ell$, gives
\[
 \E\!\left[\Phi(\widetilde G)\mid
   \widetilde g_i=g_i\text{ for }1\le i<\ell\right]
 = \Phi\!\left(\mathbf w(g_1),\ldots,\mathbf w(g_{\ell-1}),
               \bar{\mathbf w}_\ell,\ldots,\bar{\mathbf w}_q\right).
\]
The right-hand side is exactly the value of $\Phi(x)$ before \textsc{Select}
processes $B_\ell$. By~\eqref{eq:select-bucket-mean}, the trial values
obtained by trying each good in $B_\ell$ average to this current value.
\textsc{Select} chooses a minimizer, so the new value is at most the
current one. Initially all columns hold their means, giving
$\E[\Phi(\widetilde G)]$. After the last choice all columns hold the
selected weights, giving $\Phi(G)$. Hence $\Phi(G)\le\E[\Phi(\widetilde G)]$
by the method of conditional expectations~\cite{Raghavan}.

It remains to check the two objectives used by the algorithm. Each term of
$Q_t$ uses two distinct buckets, and only one factor in $P_t^+$ depends on
column $\ell$. Thus $Q_t$ and $P_t^+$ are affine in that column, as are
their weighted sums $\Phi_1$ and $\Phi_2$. For either objective, maintaining
each type's row sum and product gives the comparison coefficients in $O(k)$
operations per bucket (Appendix~\ref{app:general-cost}). Each good in a
full bucket is then evaluated once in $O(k)$ operations. Including the
initial computation of the bucket means, the total is $O(k(m+1))$.
\end{proof}

\subsection{The first sample}\label{sec:first-sample}
Before the first sample, Lemma~\ref{lem:trim} gives $D_t=0$ and $w_t(R)=N$.
If $r=2$, the next agent receives the remainder, so removing at most
$2-\beta$ for each type leaves her at least $\beta$. At the same time, the
products must be small enough to start the analysis of later samples.

\begin{lemma}[First-sample deficit]\label{lem:first-deficit}
Immediately after the first sample, $\sum_{t\in K_+}P_t\le h^3/64$.
At the start of every round with $s=1$ and for each $t\in K_+$,
\begin{equation}\label{eq:first-deficit}
 D_t\le 2-\beta .
\end{equation}
\end{lemma}
\begin{proof}
Fix $t\in K_+$ and write $p_{t\ell}=\E[w_t(\widetilde g_\ell)]=w_t(B_\ell)/H$.
Since $H=hN$, their sum is at most $w_t(R)/(hN)=1/h$.
Here $P_t=1$ because no sample has yet been taken. Independence gives
\[
 \E[Q_t(\widetilde G)]=\sum_{\ell\ne\ell'}p_{t\ell}p_{t\ell'}\le h^{-2},
 \qquad
 \E[P_t^+(\widetilde G)]=\prod_\ell(1+(h-1)p_{t\ell})
       \le e^{1-1/h}<3 .
\]
For any candidate $G$, set $x=x(G)$. Initial weights lie in $[0,1]$ and
trimming only lowers them toward zero, so $0\le x_{t\ell}\le1$.
The degree-two Bonferroni inequality gives
\[
 0\le\prod_\ell(1-x_{t\ell})\le 1-\sum_\ell x_{t\ell}+
       \sum_{\ell<\ell'}x_{t\ell}x_{t\ell'}=1-w_t(G)+Q_t(G)/2 .
\]
Thus $Q_t(G)\le2-1/h$ suffices to keep $w_t(G)\le2-\beta$.

Let $G$ be the actual choice. The objective sums over $|K_+|\le k$ types.
Since $h-212=4\lceil\sqrt{k/32}\rceil\ge\sqrt{k/2}$, we have
$k\le2(h-212)^2$. Lemma~\ref{lem:select} now gives
\[
 \Phi_1(G)\le \E[\Phi_1(\widetilde G)]\le\frac{k}{h^2}(1+384/h)
     \le 2(1-212/h)^2(1+384/h)\le 2\,\frac{h+384}{h+424}
     \le 2-\frac1h .
\]
Indeed, for $0\le x\le1$ we have $(1-x)^2(1+2x)=1-3x^2+2x^3\le1$, so
\[
 (1-212/h)^2\le \frac{h}{h+424}.
\]
The offset $212$ is a convenient multiple of four that satisfies $2\cdot212>384$
and gives a minimum width parameter $h=216>e^{16/3}$, as needed below.
The last inequality uses $80h\ge h+424$. By nonnegativity,
$Q_t(G)\le2-1/h$ for every protected type. The coefficient $128/h^3$ in
$\Phi_1$ also gives $\sum_{t\in K_+}P_t^+(G)\le 2h^3/128=h^3/64$.
The deficit rises by $w_t(G)$ in this round and cannot increase again
while $s=1$.
\end{proof}

\paragraph{Using the product alone at the first sample.}
Let $E$ be the event that some protected type loses more than $2-\beta$
under the random draw $\widetilde G$. Since $h^{w_t(\widetilde G)}\le
P_t^+(\widetilde G)$, Markov's inequality and a union bound give
$\Pr(E)\le k e^{1-1/h}/h^{2-\beta}=(e+o(1))k/h^2$. This bound crosses
one at $h\sim\sqrt{ek}$, where the coefficient of $1/\sqrt{k}$ in
$\beta=1/(2h)$ is $1/(2\sqrt e)\approx0.30$. Using $Q_t(\widetilde G)/2$
gives $\Pr(E)\le k/[2(1-\beta)h^2]$, which allows $h\sim\sqrt{k/2}$ and
yields the coefficient $1/\sqrt2$.

\subsection{Later samples}\label{sec:later-samples}
\begin{claim}[Cumulative product]\label{cl:general-product}
At the start of every round with $s\ge1$,
$\sum_{t\in K_+}P_t\le(h^3/64)e^{4(s-1)}$.
\end{claim}
\begin{proof}
Lemma~\ref{lem:first-deficit} supplies the bound at $s=1$. In a later
sampling round $H=bN$. By Lemma~\ref{lem:trim}, $w_t(R)\le N$, so
$\sum_\ell w_t(B_\ell)/H\le w_t(R)/(bN)\le 1/b$.
The current product $P_t$ is fixed, and independence gives
\[
 \E[P_t^+(\widetilde G)]\le P_t\exp((h-1)/b)\le e^4 P_t
\]
because $b=h/4$. Sum over $K_+$ and apply Lemma~\ref{lem:select} to
$\Phi_2$. Each sample multiplies the bound on the sum by at most $e^4$,
and empty or singleton rounds preserve the products.
\end{proof}

\begin{lemma}[Later-sample deficit]\label{lem:later-deficit}
At the start of every round with $s\ge2$, each $t\in K_+$ satisfies
\begin{equation}\label{eq:later-deficit}
 D_t\le3-\beta+\tfrac34(s-2).
\end{equation}
\end{lemma}
\begin{proof}
By Lemma~\ref{lem:trim} and Claim~\ref{cl:general-product},
$h^{D_t}\le P_t\le(h^3/64)e^{4(s-1)}$. Taking logarithms gives
\[
 D_t\le 3-\frac{\ln64-4}{\ln h}+(s-2)\frac4{\ln h}.
\]
At $s=2$ the margin is at least $\beta$, since $\ln64-4>1/8$ and
$\ln h/(2h)\le1/8$ for $h\ge216$. Each further sample adds $4/\ln h\le3/4$
to this upper bound because $h\ge216>e^{16/3}$.
\end{proof}

\subsection{Proof of the value guarantee}
\begin{lemma}[Weight available to each agent]\label{lem:every-payment}
At every nonfinal arrival, $w_t(R)\ge N/4+\beta$ for all $t\in K_+$.
At the final arrival, $w_t(R)\ge\beta$ for all such types.
At every arrival with $t_j\in K_+$, $w_{t_j}(T_j)\ge\beta$.
\end{lemma}
\begin{proof}
At a nonfinal arrival $N=r+s\ge s+2$. For $s=0$, Lemma~\ref{lem:trim}
gives $D_t=0$ and hence $w_t(R)=N$.
For $s=1$, $N\ge3$ and~\eqref{eq:first-deficit} gives
$w_t(R)\ge N-2+\beta\ge N/4+\beta$. For $s\ge2$, \eqref{eq:later-deficit}
gives $D_t\le3(s+2)/4-\beta\le3N/4-\beta$.
At the final arrival $N=1+s$. The same three cases give $w_t(R)=1$,
$w_t(R)\ge\beta$, and $w_t(R)\ge\beta+(s-2)/4$, respectively.

A positive-TPS agent who receives the remainder or a qualifying singleton
therefore meets her weight target. In a sampling round all of her item
weights are below $\beta$. At the first sample this gives
$|R|>N/\beta=2hN$ and at any later sample $|R|>(N/4)/\beta=2bN$.
Thus full buckets exist in both cases. Lemma~\ref{lem:payment} gives
each candidate a weight of at least $1/h-\beta=\beta$ at the first sample
and at least $1/(4b)-\beta=\beta$ thereafter.
\end{proof}

\begin{proof}[Proof of Theorem~\ref{thm:general}]
Fix any instance and type sequence. For every $j$ with $t_j\in K_+$,
Lemmas~\ref{lem:trim} and~\ref{lem:every-payment} give
$v_{t_j}(T_j)\ge d_{t_j}w_{t_j}(T_j)\ge d_{t_j}/(2h)$.
For $d_{t_j}=0$ the target follows from nonnegativity.
Taking the worst case gives $\alpha_k^{\mathrm{TPS}}\ge1/(2h)$.
Section~\ref{sec:model} gives $\alpha_k\ge\alpha_k^{\mathrm{TPS}}$, and
$h\le\sqrt{k/2}+216$ gives $1/(2h)\ge1/(\sqrt{2k}+432)$.
The limit inferior follows from $h=\sqrt{k/2}+O(1)$.
Lemma~\ref{lem:general-cost} establishes the running time.
\end{proof}

\subsection{Running time}
\begin{lemma}[Operation count and bit complexity]\label{lem:general-cost}
Algorithm~\ref{alg:general} uses $O(k(m+1)\log(m+2))$ arithmetic
operations and comparisons in preprocessing and $O(k(m+1)+m\log(m+2))$
per arrival. The operands have $O(k(m+1)L)$ bits.
\end{lemma}
\begin{proof}
Sort the item values once for each type and evaluate~\eqref{eq:tps-closed}
with suffix sums. On an arrival, sorting the goods for the revealed type
costs $O(m\log(m+2))$. Scanning or trimming all types' weights and running
\textsc{Select} cost $O(k(m+1))$ by Lemma~\ref{lem:select}.
Appendix~\ref{app:general-cost} gives the arithmetic updates and proves
the bit bound.
\end{proof}

\section{Binary valuations with two arrivals}\label{sec:binary}
A \emph{binary} type approves a subset of the goods and assigns value one
to each approved good and zero to each other good. A type approving $a$
goods has two-way MMS $\lfloor a/2\rfloor$. Let
\[
 \mathcal I^{\mathrm{bin}}_{k,2}
   =\{I\in\mathcal I_{k,2}:v_t(g)\in\{0,1\}\text{ for all }t,g\}.
\]
Define $\alpha^{\mathrm{bin}}_{k,2}$ by restricting the infimum in the
definition of $\alpha_{k,2}$ to $\mathcal I^{\mathrm{bin}}_{k,2}$.
For $p\ge2$, the pair construction on $2p$ goods uses $1+\binom{2p}{2}$ types.
For each $k\ge1$, let
\[
 p_k=\max\left\{p\in\mathbb Z_{\ge0}:1+\binom{2p}{2}\le k\right\}
     =\left\lfloor\frac{1+\sqrt{8k-7}}{4}\right\rfloor .
\]

\begin{proposition}[Counting the pair obstruction of Kulkarni et al.]\label{prop:upper}
For $k\ge7$,
$\alpha_{k,2}^{\mathrm{TPS}}\le\alpha_{k,2}\le\alpha^{\mathrm{bin}}_{k,2}\le 1/p_k$.
In particular, $\limsup_{k\to\infty}\sqrt{k}\,\alpha_k\le\sqrt 2$.
\end{proposition}
\begin{proof}
Kulkarni et al.~\cite[Theorem~3.3 and Section~4.2]{KMS} give the
obstruction. Specializing it to two arrivals and counting its types
exactly gives the stated dependence on $k$. There are $2p_k$ goods. One type approves all goods. For each unordered
pair, one type approves exactly the two goods in that pair. Any additional
listed types may remain absent. Send the all-goods type first. Its share
is $p_k$, so a bundle of size at most one yields a fraction at most $1/p_k$.
If the bundle contains two or more goods, send a type supported on a pair
in that bundle. This agent has share one and receives value zero. A binary
instance belongs to the general class, and TPS dominates MMS. Finally
$p_k=\sqrt{k/2}+O(1)$.
\end{proof}

\begin{theorem}\label{thm:binary}
For every $k\ge1$, $\alpha^{\mathrm{bin}}_{k,2}=1/\max\{1,p_k\}$, attained
by a deterministic polynomial-time policy.
In particular, $\alpha^{\mathrm{bin}}_{k,2}=1$ exactly when $k\le6$, and
$\lim_{k\to\infty}\sqrt{k}\,\alpha^{\mathrm{bin}}_{k,2}=\sqrt 2$.
\end{theorem}
\begin{proof}
Let $M_t=\{g\in M:v_t(g)=1\}$ be the approval set of type $t$. Fix the first
type $A$ and put $m_A=|M_A|$. We choose $T_1\subseteq M_A$ for the first agent
and give $M\setminus T_1$ to the second. For $k=1$, split the approved goods as evenly as possible.

\emph{Two to six types.}
Here $p_k=1$, so both agents need their full MMS. Index the other types by
distinct elements of $[5]$. The \emph{pattern} of $g\in M_A$ is the set of these
types that approve $g$. Whenever two goods have patterns $X\subseteq Y$ (possibly
equal), put the good with pattern $X$ in $T_1$ and reserve its partner. Every
type approving the selected good also approves the reserved one.

The unpaired goods have distinct patterns forming an antichain $\mathcal F$.
Lemma~\ref{lem:five-patterns} selects $\lfloor|\mathcal F|/2\rfloor$ of them for $T_1$.
In each coordinate the selected count is at most one plus the unselected count.
Adding back the pairs preserves this bound, and all goods outside $M_A$ remain. Thus for every $t\ne A$,
\[
 |M_t\cap T_1|\le |M_t\setminus T_1|+1,
 \qquad v_t(M\setminus T_1)=|M_t\setminus T_1|\ge\lfloor |M_t|/2\rfloor=\mu_t .
\]
Since $|T_1|=\lfloor m_A/2\rfloor$, type $A$ receives full MMS in either position.
Repeated scans find comparable pairs in polynomial time, and the residual search has constant size.

\emph{Seven or more types.}
Here $p_k\ge2$. Proposition~\ref{prop:upper} gives the upper bound. For each type $t$, let $a_t$ be its number of approved goods and $\nu_t=\lceil\lfloor a_t/2\rfloor/p_k\rceil$ its required number. Give $A$ exactly $\nu_A$ approved goods. If $\nu_A\le1$, any such choice leaves each other type at least its required number. A second arrival of type $A$ is covered since $m_A\ge2\nu_A$.

Suppose $\nu_A\ge2$. Every type $t\ne A$ with $\nu_t=0$ or $a_t-\nu_A\ge\nu_t$ already has enough goods after any first choice of this size. For each remaining type, designate $\nu_t$ disjoint pairs of its approved goods. This is possible because $\nu_t\le\lfloor a_t/2\rfloor$. Build a simple graph on $M_A$ by adding every designated pair whose two endpoints lie in $M_A$. An independent first bundle leaves at least one good in each designated pair, including pairs with a good outside $M_A$.

For a type contributing edges, $a_t\le\nu_A+\nu_t-1$, while the definition of $\nu_t$ gives $2p_k\nu_t\le a_t+2p_k-2$. Hence $\nu_t\le1+(\nu_A-2)/(2p_k-1)$. Each such type contributes at most $\nu_t$ edges. There are at most $k-1$ such types, so the number $E$ of distinct edges satisfies
\[
 E\le(k-1)\left(1+\frac{\nu_A-2}{2p_k-1}\right)
   \le\left(\binom{2p_k+2}{2}-1\right)\left(1+\frac{\nu_A-2}{2p_k-1}\right),
\]
where the last inequality uses the maximality of $p_k$. The graph has $m_A\ge2p_k(\nu_A-1)+2$ vertices by the definition of $\nu_A$. Lemma~\ref{lem:independent-bundle} with $p=p_k$ therefore finds an independent set of size $\nu_A$ in polynomial time. Fix the order of all choices. Finally $p_k=\sqrt{k/2}+O(1)$ gives the stated limit.
\end{proof}

\begin{lemma}[Five approval coordinates]\label{lem:five-patterns}
Every antichain $\mathcal F\subseteq 2^{[5]}$ has a subfamily $\mathcal X$
of size $\lfloor|\mathcal F|/2\rfloor$ such that, for each $i\in[5]$,
\[
  |\{X\in\mathcal X:i\in X\}|
    \ \le\ \left\lceil |\{F\in\mathcal F:i\in F\}|/2\right\rceil .
\]
\end{lemma}
\begin{proof}
Label each pattern $F\subseteq[5]$ by the integer $\sum_{i\in F}2^{i-1}$.
Starting from the empty family, append patterns in increasing label order,
trying every choice that contains no earlier member. A proper subset has
a smaller label than its superset, so a new pattern cannot be a subset
of an earlier member either. Each family reached is therefore an antichain.
Conversely, every antichain can be built in this way by appending its
members in their unique increasing order. Thus each antichain is visited
exactly once.

The Python~3 program in Appendix~\ref{app:five-patterns} visits $7{,}581$
antichains in total, including the empty family. For each $\mathcal F$, it
finds a subfamily of size $\lfloor|\mathcal F|/2\rfloor$ satisfying all
five coordinate bounds. The largest antichains have ten members. On a
given instance with $2\le k\le6$, the policy in Theorem~\ref{thm:binary}
searches only the antichain left after pairing, so it tests at most
$\binom{10}{5}=252$ candidate subfamilies.
\end{proof}

\begin{lemma}[An independent first bundle]\label{lem:independent-bundle}
Let $p\ge2$ and $\nu\ge2$ be integers. A simple graph with $m'\ge2p(\nu-1)+2$ vertices and at most
\[
 \left(\binom{2p+2}{2}-1\right)\left(1+\frac{\nu-2}{2p-1}\right)
\]
edges has an independent set of size $\nu$, computable in polynomial time.
\end{lemma}
\begin{proof}
Write $J_p=\binom{2p}{2}$ and $E$ for the edge count. For $\nu=2$ the graph has at least $2p+2$ vertices and at most $J_{p+1}-1$ edges, so a missing edge gives the desired pair.

For $\nu\ge3$, repeatedly select a vertex of minimum degree and delete its closed neighborhood. If its degree is $d$, the deleted vertices contribute at most $(d+1)/(1+d)=1$ to $\sum_v 1/(1+\deg v)$, and the surviving contributions can only increase. Thus the greedy algorithm finds an independent set of size at least this sum (the Caro--Wei bound~\cite{Caro,Wei}), which by Cauchy--Schwarz is at least $(m')^2/(m'+2E)$. This is also the average-degree bound from Tur\'an's theorem~\cite{Turan}.

Put $z=\nu-1\ge2$. If the last fraction were at most $z$, then
\[
 E\ge\left\lceil\frac{m'(m'-z)}{2z}\right\rceil\ge zJ_p+4p .
\]
The unrounded expression is increasing for $m'\ge2pz+2$ and equals $zJ_p+4p-1+2/z$ at the lower endpoint. On the other hand, $J_{p+1}-1=J_p+4p$ and $(J_p+4p)/(2p-1)<J_p$, since $(2p-1)^2>2p+3$ for $p\ge2$. The edge hypothesis therefore gives
\[
 E\le (J_p+4p)+(z-1)\frac{J_p+4p}{2p-1}<zJ_p+4p .
\]
This contradiction gives an independent set of size at least $\nu$ in at most $m'$ greedy steps.
\end{proof}

\begin{proof}[Proof of Proposition~\ref{prop:oblivious}]\phantomsection\label{proof:oblivious}
Theorem~\ref{thm:general} gives a deterministic guarantee of at least
$1/(\sqrt{2k}+432)\ge 1/(434\sqrt{k})$ at every failure probability.
For the other lower bound, fix an integer $h\ge16$ and put $\beta=1/(8h)$.
We construct a lottery that gives every agent at least $\beta$ times her
MMS with probability at least $1-15/h^2$ on each fixed sequence.

\emph{The lottery.}
Use Algorithm~\ref{alg:general}'s initial weights, its last-agent, zero-TPS,
and singleton rules, and its trimming rule, with target $\beta$. With $r$ agents
left and $s$ earlier samples, put $N=r+s$. On every sampling round use
width $H=hN$ and replace \textsc{Select} by drawing one good uniformly
from each full sorted bucket, independently across buckets. Increment $s$
even when there are no full buckets. At each node, draw a branch independently
for each possible next type and then draw its continuation from the resulting
state. Sampling this finite tree in advance gives one lottery with the
prescribed draws on each fixed sequence. Each outcome is a feasible policy
since it allocates only remaining goods and gives the last agent the remainder,
including after an earlier miss.

\emph{Remaining weight.}
Fix the sequence and a type $u$ with $d_u>0$. Put $W_u=w_u(R)$,
$D_u=N-W_u$, and $Z_u=h^{D_u}/3^s$. The weight updates in
Lemma~\ref{lem:trim} hold for every choice of representatives.
The deficit $D_u$ starts at zero, stays nonnegative, and can only decrease
on nonfinal nonsampling rounds. A sample $G$ increases it by $w_u(G)$.
Thus $W_u=r$ whenever $s=0$. Before a sample the full-bucket means
$\bar w_{uB}=w_u(B)/H$ sum to at most $W_u/H\le1/h$.
By independence and $h^a\le1+(h-1)a$ for $0\le a\le1$,
\[
 \E\!\left[h^{w_u(G)}\mid\text{history}\right]
 \le \prod_B\bigl(1+(h-1)\bar w_{uB}\bigr)
 \le \exp\!\left((h-1)\sum_B\bar w_{uB}\right)\le e<3 .
\]
Since $3^s$ triples on a sample, $\E[Z'_u\mid\text{history}]\le Z_u$
on every nonfinal round. Initially $Z_u=1$, so $\E[Z_u]\le1$ before every allocation.

\emph{Probability of a missed target.}
A type-$t$ agent who misses $\beta\mu_t$ has $d_t>0$ and receives a
bundle $G$ with $w_t(G)<\beta$, since $v_t(G)\ge d_tw_t(G)$ and
$d_t\ge\mu_t$. If $r\ge2$, the round must be a sample, so every good in
$R$ also has weight below $\beta$. Lemma~\ref{lem:payment} gives the
pre-draw bound $W_t<2\beta H=N/4$. If $r=1$, the agent gets $R$, so
$W_t<\beta$. Either miss requires $s\ge1$ because $W_t=r$ when $s=0$.

For a nonfinal miss, write $r=2+j$ and $s=1+\ell$. Then
$D_t>3(r+s)/4\ge2+(j+\ell)/2$. Since $\sqrt h\ge4$,
$h^{D_t}\ge h^2 2^j3^\ell$, so $Z_t\ge h^2 2^j/3$.
For a final miss, $h^{W_t}\le h^\beta\le e<3$ because $\log h/(8h)\le1$.
Hence $Z_t=h^2(h/3)^{s-1}/(3h^{W_t})\ge h^2/9$.
The type at each position is fixed before the draw. Markov's inequality
with $\E[Z_t]\le1$ therefore bounds the probability of a miss at $r\ge2$
by $(3/h^2)2^{-(r-2)}$, and at $r=1$ by $9/h^2$. A union bound over the
agents gives total failure probability at most
\[
 \frac9{h^2}+\frac3{h^2}\sum_{j=0}^\infty 2^{-j}=\frac{15}{h^2}.
\]
For $0<\delta<1$, take $h=\lceil16/\sqrt\delta\rceil$. Then
$16\le h\sqrt\delta\le17$, so $15/h^2\le\delta$ and
$1/(8h)\ge\sqrt\delta/136$. Together with the deterministic guarantee
this proves the lower bound.

\emph{Upper bound.}
For $k\ge7$, use the two-agent instance of Proposition~\ref{prop:upper} with $2p_k$ goods. The all-goods type arrives first and has MMS $p_k$. Each pair of goods is the support of a possible second type with MMS one.

Suppose a policy guarantees $x>0$ with failure probability at most $\delta$ on each fixed sequence. If its first bundle contains fewer than $\lceil xp_k\rceil$ goods, the first agent misses the target on every continuation. Otherwise at least $\binom{\lceil xp_k\rceil}{2}$ pairs lie inside the bundle, and their types would receive zero. The first bundle has the same distribution for every second type. Averaging the fraction of failing pairs over this distribution shows that some fixed pair has failure probability at least $\binom{\lceil xp_k\rceil}{2}/\binom{2p_k}{2}$. Hence
\[
 \delta\ge\frac{\binom{\lceil xp_k\rceil}{2}}{\binom{2p_k}{2}}
        \ge\frac{(\lceil xp_k\rceil-1)^2}{4p_k^2}.
\]
It follows that $x\le 1/p_k+2\sqrt{\delta}$. Maximality of $p_k$ gives
$k\le\binom{2p_k+2}{2}=(p_k+1)(2p_k+1)\le6p_k^2$, so
$x\le\sqrt{6/k}+2\sqrt{\delta}\le5\max\{k^{-1/2},\sqrt{\delta}\}$.
For $k\le6$, $x\le1\le\sqrt{6/k}$ gives the same upper bound.
\end{proof}

\section{What can two known types guarantee?}\label{sec:two-overview}
With two types, the current agent has just one other valuation to protect. We use this to prove $4/7\le\alpha_2\le9/10$. The lower endpoint is the guarantee of a particular policy. The upper endpoint is the exact optimum for three agents.

\begin{theorem}[Two types, any number of agents]\label{thm:two}
For two known nonnegative additive valuations and every $n\ge1$, a deterministic online policy gives every type-$t$ arrival at least $4\mu_t/7$ on every type sequence. With explicit rational values it runs in polynomial time.
\end{theorem}
\begin{proof}
Choose the scales $y_u\ge4\mu_u/7$ from Appendix~\ref{sec:score}. When $y_u>0$,
$v_u(S)\ge y_u\score_u(S)$ for every bundle $S$. When $r$ agents remain, let $R$
be the goods still available. We maintain $\score_u(R)\ge r$ for every type $u$ with $y_u>0$.
Initially this is the admissibility condition $\score_u(M)\ge n$.

Suppose $r\ge2$. A type-$t$ agent with $y_t=0$ receives the empty bundle, meeting
her zero target and leaving $R$ intact. Otherwise Lemma~\ref{lem:pack} gives $r$
disjoint candidates $G_1,\ldots,G_r\subseteq R$. Each $G_i$ is worth at least $y_t$
to her and leaves $\score_t(R\setminus G_i)\ge r-1$. Let $u$ be the other type.
If $y_u=0$, then $\mu_u=0$ and only $t$ needs the score bound. Every candidate preserves it, so give
her the first in a fixed order. If $y_u>0$, choose a candidate $G$ of least $u$-score.
The candidates' $u$-scores sum to at most $\score_u(R)$ by disjointness, so $\score_u(G)\le\score_u(R)/r$
and $\score_u(R\setminus G)\ge(1-1/r)\score_u(R)\ge r-1$.

If the last agent has type $t$, her bundle $R$ is worth at least $y_t$ by the invariant
when $y_t>0$ and by nonnegativity otherwise. This guarantees $(4\mu_t/7)_{t\in[2]}$.
With fixed tie orders, the computations in Appendix~\ref{sec:compute} take polynomial time on rational inputs.
\end{proof}

\begin{proposition}[Sharpness for the policy]\label{prop:policy}
For the policy of Theorem~\ref{thm:two} using the largest admissible scales of Appendix~\ref{sec:compute}, the infimum of its worst-case MMS fraction over all two-type instances and agent counts is $4/7$.
\end{proposition}
\begin{proof}
For each $j\ge1$, Appendix~\ref{sec:limits} constructs an instance with $n=3\cdot2^j$ agents and $n$ copies of a bundle $P_j$ worth $x(P_j)$ to $A$, where $x(P_j)\to7/4$. One extra good is worth one to $A$ and zero to $B$. When $A$ arrives first, this good alone is the unique candidate with zero $B$-score. The policy gives it to her, while the $n$ copies certify $\mu_A\ge x(P_j)$. Her fraction is at most $1/x(P_j)\to4/7$. Theorem~\ref{thm:two} gives the reverse inequality.
\end{proof}

\begin{theorem}[Two types, three agents]\label{thm:three}
For two known additive types, $\alpha_{2,3}=9/10$. If $A$ arrives first, a deterministic policy using exact MMS partitions gives every $A$ at least $\mu_A$ and every agent of the other type $B$ at least $9\mu_B/10$.
\end{theorem}
\begin{proof}
Take an MMS partition for each type. By Lemma~\ref{lem:row}, one part of $A$'s partition leaves two bundles worth at least $9\mu_B/10$ each to $B$. Give this part to the first agent. The remaining goods also contain the other two parts of $A$'s partition, each worth at least $\mu_A$ to $A$. The next arrival takes the part of her own partition that the other type values less. Either type can then take the remainder (Lemma~\ref{lem:two-arrivals}).

For the upper bound, group eight goods as follows. The second row is obtained by transposing and permuting the matrix of Feige and Norkin~\cite[Prop.~18]{FN} and omitting its zero cell.
\[
\begin{array}{@{}l@{\quad}ccc@{\qquad}ccc@{\qquad}cc@{}}
 & \multicolumn{3}{c}{I}&\multicolumn{3}{c}{J}&\multicolumn{2}{c}{K}\\
 v_A&7&1&1&3&3&3&4&5\\
 v_B&7&1&1&3&3&3&6&6
\end{array}
\]
The three groups give $\mu_A=9$. For $B$, the partition $(7+3),(1+3+6),(1+3+6)$ attains the average of ten. To exceed $9/10$, integer values require at least nine for each $A$ and ten for each $B$. Send $A$ first. On continuation $AA$, every bundle must be worth exactly nine. The seven must go with both ones, and the five with the four, so the first bundle must be $I$, $J$, or $K$. As shown in the sharpness part of Lemma~\ref{lem:row}, removing any such group leaves no two bundles worth ten each to $B$. Thus continuation $BB$ defeats the same first choice.
\end{proof}
Lemma~\ref{lem:padding} gives $\alpha_{2,n}\le9/10$ for every $n\ge3$. Table~\ref{tab:few-agents} records the corresponding bounds for two to five agents.

\section{Open questions}
Does $\lim_{k\to\infty}\sqrt{k}\,\alpha_k=\sqrt{2}$ hold for unrestricted additive valuations? The binary two-arrival class attains this coefficient. Algorithm~\ref{alg:general} loses a factor of two when it balances the singleton target $\beta$ against the first sample's guaranteed weight $1/h-\beta$. A smaller upper coefficient would require another class of hard instances.

For two types, $\alpha_2$ lies in $[4/7,9/10]$. On the sharpness instances of Appendix~\ref{sec:limits}, the rule of giving the cheapest candidate for the other type forces a singleton allocation. Can another choice preserve the other type's reserve and guarantee a fixed fraction greater than $4/7$ of MMS for any number of agents? The first open case has four agents. Does $\alpha_{2,4}<9/10$ hold? By Lemma~\ref{lem:padding}, any such upper bound would continue to hold for any larger number of agents.

For the oblivious-arrival guarantee of Proposition~\ref{prop:oblivious}, set $\delta=c/k$ with fixed $c>0$. Does $\lim_{k\to\infty}\sqrt{k}\,\alpha_k^{\mathrm{obl}}(c/k)$ exist, and if so, what is its value?

\section*{Use of generative AI}

The author used GPT-6 and GPT-6.2 Astra to explore approaches to the
allocation problem, examine assumptions and boundary cases, and draft
and revise mathematical proofs. For example, this assistance helped
clarify the proof of Lemma~\ref{lem:select}: once the choices in earlier
buckets are fixed, evaluating the objective at the mean weight vector
of each remaining bucket gives its conditional expectation. The author
used Claude Opus~5.5 (Anthropic) to organize the manuscript and audit its arguments. This
included reviewing the order in which definitions and results appear,
checking statements against their proofs, and identifying gaps in the
exposition or inconsistencies in notation.

In the OpenAI Codex environment, the author used AI assistance to draft
and edit explanatory passages and to check cited theorem numbers,
bounds, and attributions against the original papers. The author also
used this environment to develop the accompanying Lean~4 proof scripts
and to write and test the Python program for
Lemma~\ref{lem:five-patterns}. The Lean kernel checked the formal
proofs. The Python program enumerates all $7{,}581$ antichains of
subsets of a five-element set and finds, for each antichain, a
subfamily of the required size satisfying all five coordinate bounds.
The author takes responsibility for the mathematical results and
proofs, the verification code, the references, and the final text.

\bibliographystyle{plainnat}
\bibliography{online_mms_references}

\appendix
\section{Two types with any number of agents}\label{sec:score}
This appendix gives the score, scale, and candidate construction for Theorem~\ref{thm:two}, and the instances used in Proposition~\ref{prop:policy}.

\subsection{Rounding the value of a good}
Fix a type $t$ and a scale $y>0$. Put $x_g=v_t(g)/y$ for $g\in M$. The score levels are
\[
 \mathcal L_2=\{2^{-j}:j\ge1\},\qquad
 \mathcal L_3=\{1/(3\cdot2^j):j\ge0\},\qquad
 \mathcal L=\{1\}\cup\mathcal L_2\cup\mathcal L_3.
\]
Set $\sigma(0)=0$ and $\sigma(x)=\max\{u\in\mathcal L:u\le x\}$ for $x>0$.
For $P\subseteq M$, put $x(P)=\sum_{g\in P}x_g$ and $\score_{t,y}(P)=\sum_{g\in P}\sigma(x_g)$.
With $t,y$ fixed, write $\score(P)$ for this score. Rounding down gives $\score(P)\le x(P)=v_t(P)/y$.

\begin{lemma}[Score of an MMS part]\label{lem:init}
If $x(P)\ge7/4$, then $\score(P)\ge1$.
\end{lemma}
\begin{proof}
Suppose $\score(P)<1$. Each good has value below one and at most one has score
$1/2$. Below $1/2$ the ratio of consecutive levels, larger over smaller, is
either $3/2$ or $4/3$. Thus a good with positive score below $1/2$ has value strictly less than $3/2$ times its score. A zero-score good has value zero.
If there is no half-score good, $x(P)\le(3/2)\score(P)<3/2$.
If there is one, its value is below one and the other goods have total score
below $1/2$. Hence $x(P)<1+(3/2)(1/2)=7/4$.
\end{proof}

\subsection{Packing disjoint candidates}
\begin{lemma}[Packing disjoint candidates]\label{lem:pack}
For every integer $r\ge1$ and remaining set $R$ with $\score(R)\ge r$, there
are $r$ disjoint candidate bundles $G_1,\dots,G_r$ such that each satisfies
\[
 x(G_j)\ge1,\qquad \score(R\setminus G_j)\ge r-1.
\]
\end{lemma}
\begin{proof}
Start with each positive-score good in its own packet. The score of a packet is the sum of its item scores. Keep packets of score one.
Process $\mathcal L_2$ and $\mathcal L_3$ separately, starting at the smallest occupied level in each.
In $\mathcal L_2$, combine pairs of equal packets until reaching one.
In $\mathcal L_3$, combine pairs up to score $1/3$ and then
combine triples of score $1/3$ into packets of score one. Every merge
preserves score and joins disjoint goods. There are finitely many goods.

At the end, unused packets of the first family have total score strictly
below $\sum_{j\ge1}2^{-j}=1$, since each score occurs at most once and
there are finitely many occupied levels. In the second family at most two
packets have score $1/3$ and at most one has each smaller score. Their
total is strictly below $2/3+\sum_{j\ge1}(3\cdot2^j)^{-1}=1$.
The residual score is strictly below two. Since $\score(R)\ge r$,
at least $r-1$ unit packets are complete, including original goods of score one.
If at least $r$ are complete, use any $r$ of them. Otherwise use the
$r-1$ unit packets and the entire remainder. That remainder has score
$\score(R)-(r-1)\ge1$. Every candidate has value at least one because $x\ge\sigma$ item by item.
Deleting a unit packet leaves score at least $r-1$, and deleting the
remainder leaves the $r-1$ unit packets.
\end{proof}

\subsection{Computing an admissible scale}\label{sec:compute}
A scale $y>0$ is \emph{admissible} for $t$ if $\score_{t,y}(M)\ge n$.
When $\mu_t>0$, setting $y=4\mu_t/7$ gives each part of an MMS partition normalized
value at least $7/4$. By Lemma~\ref{lem:init}, each part has score at least one, so this scale is admissible.

Let $y_t$ be the largest admissible scale for $t$, or zero if none exists.
When $y_t>0$, write $\score_t=\score_{t,y_t}$. If there are fewer than $n$ positive goods,
there is no admissible scale, since each good contributes at most one.
Otherwise let $a$ be the smallest positive item value and $\bar y=v_t(M)/n$.
Scale $a$ is admissible and every admissible $y$ is at most $\bar y$, since
$\sigma(v_t(g)/y)\le v_t(g)/y$. On $[a,\bar y]$ the score sum decreases by steps.
At a breakpoint $y=v_t(g)/u$, with $v_t(g)>0$ and $u\in\mathcal L$, the good still
receives score $u$, its value on the interval immediately to the left.
Thus the largest feasible scale is found by evaluating the finitely many
breakpoints in $[a,\bar y]$ and both endpoints. Each family gives a geometric
list of breakpoints $2^jv_t(g)$ or $3\cdot2^jv_t(g)$, in addition to $v_t(g)$.
The number of relevant exponents per positive good is
$O(1+\log_2(\bar y/a))$, polynomial in the binary input length. Exact rational
comparisons, score sums, and the merges in Lemma~\ref{lem:pack} have
polynomial bit and operation complexity. When either scale is positive,
$r\le n\le |M|$. These observations prove the computational assertion of
Theorem~\ref{thm:two}.

\subsection{Sharpness for the policy}\label{sec:limits}
The cutoff $7/4$ in Lemma~\ref{lem:init} is sharp for this score function.
Fix $j\ge1$ and set $n=3\cdot2^j$ and $\zeta_j=1/n$.
Let $P_j$ contain one good of value $1-2^{-j-2}$ and $n/2-1$ goods each of value
$(3/2-2^{-j})\zeta_j$. The next level above $\zeta_j$ is $(3/2)\zeta_j$.
The bundle has score $1/2+(n/2-1)\zeta_j=1-\zeta_j<1$ and value
\[
 x(P_j)=1-2^{-j-2}+(n/2-1)(3/2-2^{-j})\zeta_j
       =\frac74-5\cdot2^{-j-2}+\frac{1}{3\cdot4^j}\longrightarrow\frac74.
\]

The $n$-agent instance consists of $n$ disjoint copies of $P_j$ and one extra good.
Type $A$ values each copy as above and the extra good at one. The other type $B$ values the extra
good at zero and every ordinary good at one. At value scale one,
the $A$-score is $n(1-\zeta_j)+1=n$. At every larger scale the extra
good loses score and no good gains any, so the implemented $A$-scale
is exactly one. Type $B$ has positive scale, since it approves at
least $n$ goods.

The extra good is an original unit packet. Across the $n$ copies, the $n$ half-score goods produce $n/2$ unit packets. There are $n(n/2-1)$ goods of score $1/n$. A group of $2^j$ such goods forms a packet of score $1/3$, and three such packets form a unit. They therefore produce $n/2-1$ more unit packets. Including the extra good gives exactly $n$ unit packets, all of which are offered.

Send $A$ first. The singleton extra good is the unique candidate with zero
$B$-score, so the algorithm gives it to the first arrival. Meanwhile
$\mu_A\ge x(P_j)$ by using the $n$ copies and adding the extra good to
one of them. The first ratio is at most $1/x(P_j)\to4/7$.

\paragraph{A simpler score rule.}
The two families of levels allow a target above $6/11$. To see the limit of three score ranges, measure value in units of MMS. For a target $q\in(0,1)$, give a good of value $x$ score $\eta x$ if $x<q/2$, score $1/2$ if $q/2\le x<q$, and score one otherwise. Consider a good worth $q-\varepsilon$ together with tiny goods worth $1-q+\varepsilon$. If every bundle worth one must have a score of at least one, this example forces $\eta\ge1/[2(1-q)]$ as $\varepsilon\downarrow0$. If $q>6/11$, then $\eta q>3/5$. One good worth $q/2$ and five goods each worth $3/(10\eta)$ now have a total score of two. A candidate leaving score at least one can spend at most one. With the medium good it can afford at most one light good and falls short of $q$. Without it the candidate needs at least three light goods, so two disjoint candidates cannot both meet the target.

The endpoint $q=6/11$ works with $\eta=11/10$. Multiply values by eleven and scores by ten. Light goods now score their value below three, medium goods score five for values in $[3,6)$, and large goods score ten. A bundle worth at least eleven scores at least ten. One large or two medium goods suffice, and in every other case its score exceeds its value minus one. For packing, start with score $W\ge10r$. Use large singletons and pairs of medium goods first, each costing ten for value at least six. If $r'\ge1$ candidates are still needed, at most one medium remains. For $r'=1$ take the remainder, whose score is exactly $W-10(r-1)$ and whose value is at least eight. For $r'\ge2$ the light goods alone weigh at least $10r'-5$. Form candidates greedily to value six. Each costs less than nine, and $9(r'-1)+6\le10r'-5$ ensures enough for the last one. Deleting any candidate leaves score at least $10(r-1)$.

\section{Residual partitions for two and three agents}\label{sec:three}

\begin{lemma}[Two arrivals]\label{lem:two-arrivals}
Suppose two agents remain and, for each of the two types \(t\), the
remaining goods have a two-part partition whose parts each have value
at least \(\lambda_t\) to \(t\). Given these partitions, a deterministic
online rule gives each agent of type \(t\) value at least \(\lambda_t\).
For two agents, taking an MMS partition for each type gives \(\alpha_{2,2}=1\).
\end{lemma}
\begin{proof}
Let \(X\) be the set of goods at the start of these two arrivals and
let the first agent have type \(A\). Of the two parts in her partition,
give her the one \(B\) values less, breaking ties in a fixed order.
Each part is worth at least \(\lambda_A\) to \(A\). The part left for the
second agent also has \(B\)-value at least
\(v_B(X)/2\ge\lambda_B\) by \(B\)'s partition.
\end{proof}

When the columns attain the three-way MMS $\mu$ of $v$, the $9\mu/10$ guarantee follows from Feige and Norkin~\cite[Theorem~17]{FN}. We prove the version needed in Appendix~\ref{sec:four} for one valuation and two supplied partitions, with $\lambda$ any lower bound on the three column values.

\begin{lemma}[Choosing a part]\label{lem:row}
Let \(P_1,P_2,P_3\) be any partition of the goods into rows. Suppose a
second partition into columns \(C_1,C_2,C_3\) has \(v(C_j)\ge\lambda\)
for each \(j\), under a nonnegative additive valuation \(v\). For
some \(i\), the six remaining row-column intersections can be grouped into two bundles
that partition the complement of \(P_i\) and are each worth at least \(9\lambda/10\).
The factor \(9/10\) is tight even for an MMS column partition.
\end{lemma}
\begin{proof}
\emph{Normalize the columns.}
The lower bound is immediate for \(\lambda\le0\). Otherwise give each
good \(g\in C_j\) weight \(10v(g)/v(C_j)\). Each column weighs ten.
Because \(v(C_j)\ge\lambda\), each unit of weight contributes value
at least \(\lambda/10\). It suffices to find two bundles of weight at
least nine. Treat each cell \(P_i\cap C_j\) as a unit, and write
\(\|P_i\|\) for the weight of row \(P_i\).

Suppose that deleting any row fails to leave two bundles of weight at least nine. We can relabel the rows so that \(\|P_1\|,\|P_2\|<9\), since otherwise we could keep two whole rows of weight at least nine. Every cell in these two rows exceeds one. Indeed, if either row \(P_i\) contained a cell of weight at most one, deleting \(P_i\) would leave at least nine in that column and at least \(20-\|P_i\|>11\) in the other two columns together. Thus each cell in \(P_3\) weighs less than eight.

\emph{Claim (greedy balancing).}
Suppose two bundles start with a weight difference of \(\delta_0\), and
cells of weights \(\xi_1,\ldots,\xi_f\) are placed in a lighter
bundle in any given order. Their final difference is at most
\[
 \max\left\{\delta_0-\sum_{j=1}^{f}\xi_j,\
       \max_{1\le i\le f}\left(\xi_i-\sum_{j=i+1}^{f}\xi_j\right)\right\}.
\]
Indeed, if no cell exceeds the difference just before its placement,
each placement subtracts its weight from the difference. Otherwise,
let \(i\) be the index of the last cell whose weight exceeds the current difference just before placement.
The new difference is at most \(\xi_i\), and each later placement
subtracts its weight. This proves the claim.

\emph{Locate two large cells.}
Delete \(P_1\). Place the cells of \(P_3\) one at a time into the lighter of two initially empty bundles. The balancing claim bounds the difference \(\delta_0\) after placing \(P_3\) by its largest cell, hence \(\delta_0<8\). Next place the cells of \(P_2\) by the same rule, in decreasing order of weight \(\gamma_1\ge\gamma_2\ge\gamma_3\). Each exceeds one and their sum is less than nine, giving \(\gamma_1<7\), \(\gamma_2<4\), and \(\gamma_3<3\). The claim bounds the final difference by
\begin{equation}\label{eq:rowdiff}
 \max\{\delta_0-\|P_2\|,\ \gamma_1-\gamma_2-\gamma_3,\ \gamma_2-\gamma_3,\ \gamma_3\}<5.
\end{equation}
If \(\|P_1\|\le7\), the residual weight is at least 23, so the lighter
bundle weighs more than \((23-5)/2=9\). Hence \(\|P_1\|>7\).
Deleting \(P_2\) and placing \(P_3\) followed by the sorted cells of
\(P_1\) gives the identical bound on the difference, so
\(\|P_2\|>7\) as well. Now return to the deletion of \(P_1\). If
\(\gamma_1\le5\), the four terms in~\eqref{eq:rowdiff} are less than
\(8-7,\ 5-1-1,\ 4-1,\ 3\), respectively. The residual weight
exceeds 21, so the lighter bundle again weighs more than nine. Thus
\(P_2\) has a cell larger than five. For the deletion of \(P_2\),
the same four bounds hold if the largest cell of \(P_1\) is at most
five, giving the same contradiction. Thus \(P_1\) also has a cell
larger than five.

\emph{Split the six remaining cells.}
The two cells larger than five occupy different columns. After we permute the columns, the cell weights lie in the following intervals.
\[
\begin{array}{c|ccc}
      & C_1 & C_2 & C_3 \\ \hline
 P_1  & (5,7) & (1,3) & (1,3) \\
 P_2  & (1,3) & (5,7) & (1,3) \\
 P_3  & [0,4) & [0,4) & (4,8)
\end{array}
\]
Indeed, each of the first two rows weighs less than nine and each of its cells exceeds one. The bounds for \(P_3\) then follow from the column sums.

Delete \(P_1\). Start one bundle with the \((5,7)\) cell of \(P_2\) and the other with the \((4,8)\) cell of \(P_3\), so the initial difference is less than three. Place the two remaining cells of \(P_3\) into the lighter bundle in turn. Each weighs less than four, leaving a difference \(\delta<4\). Finally place the two remaining cells of \(P_2\), of weights \(\gamma_2,\gamma_3\in(1,3)\), by the same rule. The final difference is at most \(\max\{\delta-\gamma_2-\gamma_3,\gamma_2-\gamma_3,\gamma_3\}<3\). The residual weight exceeds 21, so both bundles weigh more than nine. This contradicts the assumed failure of every deletion.

For sharpness, take eight goods having the nonzero values in the rows
\((7,1,1)\), \((3,3,3)\), and \((0,6,6)\) obtained by transposing and permuting the example of Feige and
Norkin~\cite[Prop.~18]{FN}. The column sums and the total give
\(\mu=10\). Deleting the first row leaves value 21 in multiples of three,
so two bundles worth at least ten would require total value at least 24.
Deleting the second leaves \(7,1,1,6,6\), which have a total of 21 and no
subset of value ten or eleven. Deleting the third leaves total value 18.
Hence no deletion preserves two bundles of value strictly greater than
nine. Keeping the first two rows attains nine.
\end{proof}

\section{Alternative guarantees for finitely many types}\label{app:finite}
Algorithm~\ref{alg:general} gives the stronger
asymptotic coefficient. A single width multiplier yields a complementary
finite guarantee. Since $k$ is known, the allocator can compute both
guarantees in advance and run the algorithm with the larger one.

\begin{proposition}[One width multiplier]\label{prop:finite}
Let $h_{\mathrm f}=\max\{4,\lceil2\sqrt{k}\rceil\}$. A deterministic algorithm
guarantees each type-$t$ agent $d_t/(6h_{\mathrm f})$ for any number of agents. It takes
polynomial time on explicit rational valuations.
\end{proposition}
\begin{proof}
Use the capped weights, trimming rule, and final allocation of
Section~\ref{sec:general}. Set the target to $\beta_{\mathrm f}=1/(6h_{\mathrm f})$
and the bucket width to $h_{\mathrm f}N$ in every sampling round. Write $\gamma=3-2/h_{\mathrm f}$.
For each positive-TPS type $t$, initialize $P_t=1$ and multiply $P_t$
by $1+(h_{\mathrm f}-1)w_t(g)$ for each sampled good $g$, using its weight immediately before allocation. Choose a candidate by calling \textsc{Select} on the objective $\sum_{t\in K_+}P_t^+(x)$ of Algorithm~\ref{alg:general}, using $h_{\mathrm f}$ in the product factors.

For the analysis, let $\widetilde G$ consist of one independently and uniformly drawn good $\widetilde g_\ell$ per full bucket. Each protected type has $\sum_\ell\E[w_t(\widetilde g_\ell)]\le w_t(R)/(h_{\mathrm f}N)\le1/h_{\mathrm f}$ by the trimming rule. Hence
\[
 \E[P_t^+(\widetilde G)/P_t]
 \le \exp\!\left((h_{\mathrm f}-1)\sum_\ell\E[w_t(\widetilde g_\ell)]\right)
 \le e^{1-1/h_{\mathrm f}}\le 1+2(1-1/h_{\mathrm f})=\gamma .
\]
The last step uses the chord bound for the exponential on $[0,1]$ and $e<3$. Lemma~\ref{lem:select} therefore gives $\sum_t P_t\le k\gamma^s$ after $s$ samples. Empty and singleton rounds leave the products unchanged. By~\eqref{eq:singleton}, they also cannot increase a deficit. Since $h_{\mathrm f}^u\le1+(h_{\mathrm f}-1)u$ on $[0,1]$, the product dominates $h_{\mathrm f}^{D_t}$ as in Lemma~\ref{lem:trim}. Thus
\[
 D_t\ln h_{\mathrm f}\le\ln P_t\le\ln k+s\ln\gamma .
\]
For $s=0$ every deficit is zero. For $s\ge1$, the inequalities
$h_{\mathrm f}^2\ge4k$, $\gamma<3$, and $\gamma^3\le h_{\mathrm f}^2$ give
\[
 D_t\le \frac{\ln(k\gamma)}{\ln h_{\mathrm f}}+(s-1)\frac{\ln \gamma}{\ln h_{\mathrm f}}
       \le 2+\tfrac23(s-1).
\]
For the third inequality on the parameters, check $h_{\mathrm f}=4,5$ directly and
use $\gamma^3<27<h_{\mathrm f}^2$ for $h_{\mathrm f}\ge6$. At a nonfinal arrival, $N=r+s\ge 2+s$.
Hence $w_t(R)\ge N/3$ for every protected type $t$. In a sampling round all current-type item weights are below $\beta_{\mathrm f}$. Counting as in Lemma~\ref{lem:every-payment} gives $|R|>(N/3)/\beta_{\mathrm f}=2h_{\mathrm f}N$, so full buckets exist. Lemma~\ref{lem:payment} gives $w_{t_j}(G)\ge 1/(3h_{\mathrm f})-\beta_{\mathrm f}=\beta_{\mathrm f}$ for every candidate $G$.

At the final arrival $N=1+s$. For $s=0$ we have $w_t(R)=1$. For $s\ge1$
the product bound yields
\[
 w_t(R)\ge \frac{\ln(h_{\mathrm f}^2/k)-\ln \gamma}{\ln h_{\mathrm f}}
 +(s-1)\left(1-\frac{\ln \gamma}{\ln h_{\mathrm f}}\right)
 \ge \frac{\ln(4/3)}{\ln h_{\mathrm f}}>\beta_{\mathrm f}.
\]
Here $\ln(4/3)\ge1/4$ and $\ln h_{\mathrm f}/(6h_{\mathrm f})<1/6$. All arrivals therefore
receive the claimed fraction of TPS. The conditional means of products
are computed exactly from bucket means as in Section~\ref{sec:general},
with the same polynomial bit complexity.
\end{proof}
For example, at $k=256$ the bound with one width multiplier is $1/192$ and
Algorithm~\ref{alg:general} guarantees $1/448$. Both apply to the original TPS.

\begin{corollary}[Finite-$k$ guarantees]\label{cor:finite-envelope}
For $k\ge1$, put $h=212+4\lceil\sqrt{k/32}\rceil$ and
$h_{\mathrm f}=\max\{4,\lceil2\sqrt{k}\rceil\}$. The optimal fraction of original MMS satisfies
\[
 \alpha_k\ \ge\ L(k):=\max\{1/k,\,1/(6h_{\mathrm f}),\,1/(2h)\},
 \qquad \alpha_k>\frac{1}{\sqrt{2k}+432}.
\]
Among these three guarantees, $1/k$ is largest for $1\le k\le150$,
$1/(6h_{\mathrm f})$ for $151\le k\le1640$, and $1/(2h)$ for $k\ge1641$,
allowing ties. The values of $\sqrt{k}\,L(k)$ at $k=10^4$ and $10^6$
are approximately $0.1761$ and $0.5435$, respectively.
\end{corollary}
\begin{proof}
The first term is the MMS guarantee of Kulkarni et al.~\cite[Theorem~3.1]{KMS}.
The other two follow from Proposition~\ref{prop:finite} and Theorem~\ref{thm:general},
since $d_t\ge\mu_t$. Rounding $h$ gives $2h<\sqrt{2k}+432$.
For $k\le288$ we have $2h\ge432>k$, and for $289\le k\le512$ we have
$2h=456$. For $k\ge513$, $\sqrt{2k}+432<k$. Thus $1/(2h)>1/k$ exactly
when $k\ge457$.
For the comparisons, it suffices to minimize the integer denominators
$k,6h_{\mathrm f},2h$. For $k\ge1681$, $\sqrt{k}\ge41$ gives
$\sqrt{2k}+432<12\sqrt{k}\le\min\{6h_{\mathrm f},k\}$.
For $k\le144$, $k\le12\sqrt{k}\le6h_{\mathrm f}$. From $145$ to $150$, $6h_{\mathrm f}=150$.
In both ranges $k<2h$. For $151\le k\le155$ one still has $6h_{\mathrm f}=150$.
For $k\ge156$, $6h_{\mathrm f}<12\sqrt{k}+6\le k$, so it remains to compare $6h_{\mathrm f}$ and $2h$.
For $151\le k\le1521=39^2$, $(12-\sqrt2)\sqrt{k}+6<424$, giving
$6h_{\mathrm f}<424+\sqrt{2k}\le2h$. From $1522$ to $1568$, $2h=480$ and $h_{\mathrm f}\le80$.
From $1569$ to $1640$, $2h=488$ and $h_{\mathrm f}\le81$. From $1641$ to $1680$,
we still have $2h=488$, while $h_{\mathrm f}=82$.
The first two terms tie for the largest guarantee at $k=144,150$, and the
latter two tie for it at $1561\le k\le1568$. The numerical values concern
the certified lower bound $L(k)$. The optimum $\alpha_k$ can be larger.
\end{proof}

\section{Four and five arrivals}\label{sec:four}
\begin{table}[tb]
\caption{Two additive types with two to five agents. The lower bounds use exact MMS partitions. For fixed $\varepsilon>0$, approximate partitions yield polynomial-time policies attaining $(1-\varepsilon)$ times each lower bound (Corollary~\ref{cor:short-ptas}).}
\label{tab:few-agents}
\centering
\begin{tabular}{@{}lccl@{}}
\toprule
Number of agents & Lower bound & Upper bound & Result \\
\midrule
$n=2$ & $1$ & $1$ & Lem.~\ref{lem:two-arrivals} \\
$n=3$ & $9/10$ & $9/10$ & Thm.~\ref{thm:three} \\
$n=4$ & $3/4$ & $9/10$ & Thm.~\ref{thm:four}, Lem.~\ref{lem:padding} \\
$n=5$ & $27/40$ & $9/10$ & Cor.~\ref{cor:five}, Lem.~\ref{lem:padding} \\
\bottomrule
\end{tabular}
\end{table}
To apply the rules for two and three agents in Appendix~\ref{sec:three} after one or
more allocations, we keep an explicit partition for each type. On the
goods currently available, write \(\lambda_t\) for the minimum
type-\(t\) value of any part of its partition. The target at the
start is \(\lambda_t=\mu_t\) when we use exact MMS partitions.
If just one type has a positive target, give each of its arrivals a
distinct part of its partition and give the other arrivals empty
bundles. If both targets are zero, every allocation attains them.
In the following row arguments we may therefore take both targets
positive. Break ties in a fixed order.

\begin{lemma}[Partitions after removing the lightest rows]\label{lem:residual-rows}
For \(n\ge1\), let \(P_1,\ldots,P_n\) and \(C_1,\ldots,C_n\) be two
partitions of the available goods, with \(\lambda_A=\min_i v_A(P_i)\) and
\(\lambda_B=\min_i v_B(C_i)>0\). For \(g\in C_i\), set
\(w(g)=v_B(g)/v_B(C_i)\) and extend \(w\) additively. Index the rows so
that \(w(P_1)\le\cdots\le w(P_n)\).
For each integer \(0\le a<n\), let \(R_a=\bigcup_{i=a+1}^n P_i\) be the
goods left after allocating the first \(a\) rows, and put \(r=n-a\).
The retained rows partition \(R_a\) into \(r\) parts each worth at least
\(\lambda_A\) to \(A\). For \(B\), \(R_a\) admits an \(r\)-part
partition with each part worth at least \(n\lambda_B/(n+r-1)\).
When \(a=1\) and \(n\ge3\), \(B\) has such a partition with each part
worth at least \(3\lambda_B/4\).
\end{lemma}
\begin{proof}
Each column weighs one, each good weighs at most one, and
\(v_B(T)\ge\lambda_B w(T)\) for any subset \(T\). Write
\(\theta=n/(n+r-1)\).
If every retained row weighs at least \(\theta\), the rows suffice.
Otherwise each removed row weighs less than \(\theta\), so
\(w(R_a)\ge n-a\theta=(2r-1)\theta\).
Make goods of weight at least \(\theta\) singleton bundles and then
accumulate smaller goods until each bundle reaches \(\theta\). Stop
when \(r\) bundles are complete. Each completed bundle weighs at most
\(2\theta\), since \(2\theta\ge1\). Even after \(r-1\) bundles are complete, at least
\(\theta\) remains. Assign surplus goods to any bundle.

For \(a=1\) and \(n\ge3\), if all remaining rows weigh at least \(3/4\), they suffice.
Otherwise \(w(P_1)<3/4\). Column \(C_i\) loses \(w(C_i\cap P_1)\) and
retains \(C_i\cap R_1\). These losses sum to less than \(3/4\), so the third largest is less than \(1/4\). Merge the remainders of the
two columns with the greatest losses. Together they weigh more than
\(2-3/4\), and the \(n-2\) other columns each retain more than \(3/4\).
\end{proof}

\begin{theorem}\label{thm:four}
With four arrivals, name the first type \(A\) and the other \(B\).
A deterministic policy using an exact initial MMS partition for each type gives every
\(A\) at least \(9\mu_A/10\) and every \(B\) at least \(3\mu_B/4\).
\end{theorem}
\begin{proof}
We prove the same statement relative to any supplied partition
minima \(\lambda_A,\lambda_B\). Give the first \(A\) the row of least
weight in Lemma~\ref{lem:residual-rows}. Three \(A\) rows of value
at least \(\lambda_A\) remain, and the lemma supplies a three-part
\(B\) partition with each part worth at least \(3\lambda_B/4\). If the second arrival
is \(B\), use these \(B\) parts as rows and the retained \(A\) parts
as columns in Lemma~\ref{lem:row}. Its selected row is worth at least
\(3\lambda_B/4\) to \(B\). The remaining goods have two \(B\) parts at
this threshold and two \(A\) parts worth at least \(9\lambda_A/10\).
Lemma~\ref{lem:two-arrivals} gives these values to the final arrivals.

If the second arrival is \(A\), give it the second row in the original
order. The other two rows are each worth at least \(\lambda_A\) to \(A\).
Lemma~\ref{lem:residual-rows} at \(n=4,r=2\) supplies two \(B\)
bundles worth at least \(4\lambda_B/5\) each. Again
Lemma~\ref{lem:two-arrivals} handles every remaining sequence.
Taking \(\lambda_t=\mu_t\) proves the theorem.
\end{proof}
Together with Lemma~\ref{lem:padding} and Theorem~\ref{thm:three},
this gives \(3/4\le\alpha_{2,4}\le9/10\).

\begin{corollary}\label{cor:five}
With five arrivals, name the first type \(A\) and the other \(B\).
A deterministic policy using an exact initial MMS partition for each type gives every
\(A\) at least \(3\mu_A/4\) and every \(B\) at least \(27\mu_B/40\).
\end{corollary}
\begin{proof}
Start from any supplied partitions with minima \(\lambda_A,\lambda_B\).
Give consecutive initial \(A\) arrivals the successive rows in the
weight order of Lemma~\ref{lem:residual-rows}. If every arrival is
\(A\), each receives at least \(\lambda_A\). Otherwise write \(r\)
for the number remaining, including the first \(B\). Each of the
\(r\) residual rows is worth at least \(\lambda_A\) to \(A\).

If \(r=4\), Lemma~\ref{lem:residual-rows} gives four \(B\) parts worth
at least \(3\lambda_B/4\) each. Use these and the residual \(A\) rows
in the four-arrival rule, with \(B\) as its first type. It guarantees every
\(B\) at least \((9/10)(3/4)\lambda_B\) and every \(A\) at least
\(3\lambda_A/4\). For \(r=3\), Lemma~\ref{lem:residual-rows} gives
three \(B\) parts worth at least \(5\lambda_B/7\) each. Using them
as rows and the retained \(A\) parts as columns, the three-arrival
rule guarantees this value to each remaining \(B\) agent and at least
\(9\lambda_A/10\) to each remaining \(A\) agent. For \(r=2\), the two \(B\) parts are each worth at
least \(5\lambda_B/6\). The two-arrival rule preserves both
types' thresholds. For \(r=1\), the final row is worth at least
\(\lambda_B\) by the same lemma. Taking \(\lambda_t=\mu_t\) gives
the stated guarantees.
\end{proof}
Thus \(27/40\le\alpha_{2,5}\le9/10\).

\subsection{Polynomial-time approximate partitions}
For the two- through five-agent policies in Table~\ref{tab:few-agents},
an exact initial MMS partition for each type can be found by exhaustive search.
\begin{corollary}\label{cor:short-ptas}
For two known additive types, \(2\le n\le5\), and any fixed
\(0<\varepsilon<1\), a deterministic policy attains
\((1-\varepsilon)\) times each of the type-specific MMS guarantees
of Lemma~\ref{lem:two-arrivals}, Theorems~\ref{thm:three} and
\ref{thm:four}, and Corollary~\ref{cor:five}, in time polynomial
in the binary input length.
\end{corollary}
\begin{proof}
For each type \(t\), view its item values as job sizes and the \(n\)
parts as identical machines. A schedule's least machine load equals
the minimum part value, and the optimum is \(\mu_t\). Woeginger's
PTAS for maximizing the minimum machine completion time~\cite{Woeginger}
returns a partition whose minimum part value satisfies
\(\lambda_t\ge(1-\varepsilon)\mu_t\). For rational inputs, clearing
denominators within each type gives integer job sizes with bit lengths
polynomial in the input. If there are fewer than \(n\) positive jobs,
\(\mu_t=0\) and any partition suffices. Otherwise jobs of size zero
may be added back to any part after scheduling the positive jobs.

Run the policy using the two supplied partitions. The two-arrival
rule uses their stated minima. The three-arrival rule obtains its
residual partitions by checking three row deletions and at most
\(2^6\) splits per deletion. At four and five arrivals,
Lemma~\ref{lem:residual-rows} constructs the next partitions by
comparing row weights, merging columns, or greedily grouping goods.
Thus every continuation uses partitions already constructed on its
remaining goods. For example, when there are five agents and the first two have types
\(A\) and \(B\), respectively, the two successive factors give
\((9/10)(3/4)\lambda_B\ge(27/40)(1-\varepsilon)\mu_B\).
Every branch has just one factor \(1-\varepsilon\), from the initial
partition for the agent's type. The subsequent constructions take
polynomial time for \(n\le5\).
\end{proof}

\section{Verification for five approval coordinates}\label{app:five-patterns}
The accompanying Python program checks all five inequalities of Lemma~\ref{lem:five-patterns} for each antichain and counts the families by size. It uses only the standard library. On an arm64 Mac with CPython~3.12.14, it verified all $7{,}581$ antichains in $0.06$ seconds (median wall time over seven runs, including interpreter startup). From the source directory, run
\begin{verbatim}
python3 manuscript-sections/verify_five_patterns.py
\end{verbatim}
\begin{verbatim}
from itertools import combinations

if not __debug__:
    raise RuntimeError('Run with assertions enabled (without -O).')

counts = [0] * 33
bits = [1, 2, 4, 8, 16]

def approvals(family, bit):
    return sum(bool(mask & bit) for mask in family)

def check(family, available):
    size = len(family)
    caps = [(approvals(family, bit) + 1) // 2 for bit in bits]
    assert any(all(approvals(part, bit) <= cap
                   for bit, cap in zip(bits, caps))
               for part in combinations(family, size // 2))
    counts[size] += 1
    for j, mask in enumerate(available):
        rest = [x for x in available[j + 1:] if (x & mask) != mask]
        check(family + [mask], rest)

check([], list(range(32)))
assert counts[:11] == [1, 32, 285, 1090, 2020, 2146, 1380, 490, 115, 20, 2]
assert sum(counts) == 7581
print(f'{sum(counts):,} antichains verified.')
\end{verbatim}

\section{Implementation of Algorithm~\ref{alg:general}}\label{app:general-cost}
We give the arithmetic and bit bounds used in Lemmas~\ref{lem:select}
and~\ref{lem:general-cost}. When $m=0$, the empty-bundle branch requires constant work per arrival and skips all type scans. For $m>0$ and each type, sort the values in decreasing
order, form their suffix sums, and minimize the at most $m+1$ ratios
in~\eqref{eq:tps-closed}. This handles $m<n$ without a
loop over $n$. The extra cost of finding the smallest integer $z$ with
$32z^2\ge k$ by binary search is $O(\log(k+1))$, within the preprocessing bound.

\paragraph{Comparing goods in one bucket.}
At any step of \textsc{Select}, the array $x$ holds the weights of the
chosen good in each fixed column and the bucket means in the remaining
columns. For each $t\in K_+$, maintain
\[
 A_t=\sum_\ell x_{t\ell},\qquad
 C_t=P_t^+(x)=P_t\prod_\ell(1+(h-1)x_{t\ell}).
\]
When fixing bucket $\ell$, put
$\gamma_{t\ell}=(h-1)C_t/(1+(h-1)x_{t\ell})$.
Up to an additive constant shared by all $g\in B_\ell$, replacing column
$\ell$ by the weight vector of $g$ gives the objective value
\[
 \begin{cases}
 \displaystyle\sum_{t\in K_+}
   \bigl[2(A_t-x_{t\ell})+128\gamma_{t\ell}/h^3\bigr]w_t(g),
       &\Phi_1,\\[3pt]
 \displaystyle\sum_{t\in K_+}\gamma_{t\ell}w_t(g),&\Phi_2 .
 \end{cases}
\]
In $Q_t$, the chosen good occurs in both orders of every pair involving
its bucket. In $P_t^+$, division by the old factor leaves the product of
the factors outside that bucket. Thus all coefficients cost $O(k)$ per
bucket, and comparing its $H$ goods costs $O(kH)$. Substituting the chosen
weight in $A_t$ and replacing the old factor in $C_t$ take $O(k)$ operations.
There are at most $m$ goods in full buckets, so forming the means,
comparing candidates, and updating the products together cost $O(k(m+1))$.

\paragraph{Exact arithmetic.}
For each type, scale its values by the product of the input denominators.
The resulting integers and their sums have $O(L)$ bits. In these units
each positive TPS equals $S_c/(n-c)$ for some minimizing $c$ in~\eqref{eq:tps-closed}.
All initial weights of that type therefore share the denominator $S_c$
of $O(L)$ bits. Trimming preserves this denominator because its cap is
an integer. Bucket means introduce at most another factor $H\le hn$.
Every stored or conditional product uses at most $m$ factors. Each good
selected in an earlier sample has been removed, and the current full
buckets are disjoint subsets of the remaining goods. Each factor has
$O(L)$ bits, so these products have $O((m+1)L)$-bit numerators and
denominators. The same bound holds for the per-type coefficients above.
Summing reduced rationals over $K_+$ gives $O(k(m+1)L)$-bit numbers.
Exact comparisons and arithmetic on these operands take polynomial bit time.

\end{document}